\documentclass[aps,pre,eprint,groupedaddress]{revtex4-2}

\usepackage{graphicx}
\usepackage{xcolor}

\begin{document}

\title{Pair approximation for the $q$-voter model with independence 
on multiplex networks}

\author{T.\ Gradowski and A.\ Krawiecki}       

\affiliation{Faculty of Physics,
Warsaw University of Technology, \\
Koszykowa 75, PL-00-662 Warsaw, Poland}

\begin{abstract}
The $q$-voter model with independence is investigated on multiplex networks with
full overlap of nodes in the layers. The layers are
various complex networks corresponding to different levels of social influence. 
Detailed studies are performed for the model on multiplex networks with two layers with identical degree distributions,
obeying the LOCAL\&AND and GLOBAL\&AND spin update rules differing by the way in which
the $q$-lobbies of neighbors within different layers exert their joint influence on the opinion of a given agent.
Homogeneous pair approximation is derived for a general case of a two-state spin model on a multiplex network
and its predictions are compared with results of mean field approximation and Monte
Carlo simulations of the above-mentioned $q$-voter model with independence for a broad range of parameters. 
As the parameter controlling the level of agents' independence is changed ferromagnetic phase transition occurs
which can be first- or second-order, depending on the size of the lobby $q$. Details of this transition, e.g., position of the
critical points, critical exponents and the width of the possible hysteresis loop,
depend on the topology and other features of the layers, in particular on the mean degree of
nodes in the layers which is directly predicted by the homogeneous pair approximation.
If the mean degree of nodes is substantially larger than the size of the $q$-lobby good agreement
is obtained between numerical results and theoretical predictions based on the
homogeneous pair approximation concerning the order and details of the ferromagnetic transition.
In the case of the model on multiplex networks with layers in the form of homogeneous Erd\H{o}s-R\'enyi and random regular graphs as well as
weakly heterogeneous scale-free networks this agreement is quantitative, while 
in the case of layers in the form of strongly heterogeneous scale-free networks it is only qualitative.
If the mean degree of nodes is small and comparable with $q$ predictions of the homogeneous pair approximation
are in general even qualitatively wrong.
\end{abstract}


\maketitle


\section{Introduction}

Studies of interacting systems on complex, possibly heterogeneous networks constitute an important area of research in contemporary
statistical physics \cite{Dorogovtsev08,Barrat08}. In particular, they comprise investigation of critical phenomena 
in nonequilibrium models for, e.g., contact processes and epidemic spreading 
\cite{Pastor15}, synchronization \cite{Arenas08} and the opinion formation \cite{Castellano09a}, where the underlying network
reflects the complexity and possible heterogeneity of social interactions \cite{Albert02,Barabasi16}. An important class
of the latter models, justified from the social point of view, comprises these in which agents possessing opinions 
represented by two-state spins placed in the nodes of the network make decisions on the
basis of the opinions of randomly chosen subsets (lobbies) of their neighbors, e.g., the voter model
\cite{Sood05,Sood08,Vazquez08,Pugliese09}, 
the $q$-voter (called also nonlinear voter) model \cite{Castellano09,Przybyla11,Timpanaro14}, various versions of
the noisy $q$-voter model, e.g., with independence or anticonformism \cite{Nyczka12,Chmiel15,Jedrzejewski17,Abramiuk19,Moretti13,Peralta18a,Peralta18,Vieira18},
and the $q$-neighbor Ising model \cite{Jedrzejewski15,Park17,Chmiel17,Chmiel18}.
The above-mentioned models differ by the opinion (spin) update rules, which leads to
different critical behavior. In particular, the noisy $q$-voter models
and the $q$-neighbor Ising model can exhibit phase transition 
from the disordered paramagnetic (PM) state to the ordered ferromagnetic (FM) state with unanimous opinion as the parameter
representing agents' uncertainity in decision making (``social temperature'') is decreased. This transition can be first- or second-order,
depending on the size of the lobby $q$ and details of the degree distribution of the network.
In the case of models on complete graphs quantitatively
correct description of such transition usually can be achieved in the framework of the mean field approximation (MFA) \cite{Nyczka12,Chmiel15,Abramiuk19,Moretti13,Peralta18,Jedrzejewski15,Park17,Chmiel17}. 
However, in the case of models on networks a more accurate pair approximation (PA) taking into account dynamical correlations between interacting
agents is necessary to capture the dependence of the critical behavior on the properties of the network, e.g., on its degree distribution  \cite{Vazquez08,Pugliese09,Jedrzejewski17,Peralta18a,Peralta18,Chmiel18,Gleeson11,Gleeson13}. 

Recently it has been realized that 
even more complex and heterogeneous structures occur frequently in social systems which has prompted interest in 
the study of interacting systems on ``networks of networks'' \cite{Boccaletti14}. In this context
much attention was devoted to multiplex networks (MNs) which consist of a fixed set of nodes connected by various
sets of edges called layers \cite{Boccaletti14,Lee14,Lee15}.
Interacting systems on MNs exhibit rich variety of collective behavior and critical phenomena. For example,
percolation transition \cite{Buldyrev10,Baxter12}, cascading failures \cite{Tan13}, 
threshold cascades \cite{Kim13,Lee14a}, diffusion processes \cite{Gomez13,Sole13}, 
epidemic spreading \cite{Wu16}, asymptotic behavior of the voter model \cite{Diakonova16}, coevolution dynamics of the nonlinear voter model \cite{Min19}
and phase transitions in the equilibrium Ising model \cite{Krawiecki17,Krawiecki17a}
and related Ashkin-Teller model \cite{Jang15} as well as in a non-equilibrium majority vote model \cite{Choi19}
were studied on MNs. Also  the $q$-voter model with independence 
\cite{Chmiel15} and the $q$-neighbor Ising model \cite{Chmiel17}
were studied on MNs with layers in the form of complete graphs, both in the MFA and by means of Monte Carlo (MC) simulations. 
For the two latter models various opinion update rules 
were assumed as generalizations of the rules for the corresponding models on (monoplex) networks in order to take into account 
that agents interact with separately chosen lobbies within different layers, which correspond to different
levels of social influence. It was shown that the effect of network multiplicity on the critical
behavior (e.g., on the range of parameters for the occurrence of the discontinuous FM transition) depends on the model and on the 
assumed opinion update rule and is quantitatively well described in the MFA.

The aim of this paper is to investigate the $q$-voter model with independence, 
which is a particular kind of a general noisy $q$-voter model,
on MNs with layers in the form of complex networks rather than complete graphs.
This requires going beyond the MFA. Thus, in this paper PA is extended to a general case of two-state spin systems with
up-down symmetry on MNs with layers in the form of complex networks and with
various spin update rules. This PA is applied to the $q$-voter model with independence and the resulting
theoretical predictions concerning the FM transition are compared with those based on the simple MFA and
with results of MC simulations of the model
on MNs with layers with different degree distributions, such as random regular graphs (RRGs),
Erd\H{o}s-R\'enyi graphs (ERGs) \cite{Albert02,Barabasi16,Erdos59}
and scale-free (SF) networks \cite{Albert02,Barabasi16,Barabasi99}. For the sake of brevity the problem
under study is considered here under several simplifying assumptions. First, as natural generalizations of
the opinion update rule for the model on monoplex networks, the corresponding rules for the model on
MNs are assumed to have different AND forms \cite{Lee14a,Chmiel15,Chmiel17}: 
the agent changes opinion if interaction with every lobby from every layer suggest change
(LOCAL\&AND rule), or interaction with a set of all neighbors from all lobbies from all layers suggests change (GLOBAL\&AND rule). 
Then, only the simplest homogeneous PA is considered in which all nodes and edges
are eventually treated as equivalent and their possible heterogeneity is neglected \cite{Vazquez08,Jedrzejewski17,Peralta18a,Peralta18,Chmiel18}.
Moreover, the interest is focused only on the stationary states of the model corresponding to different (PM or FM) thermodynamic phases and their stability,
thus detailed study of the fluctuations of the macroscopic quantities characterizing the model, such as the magnetization, and their theoretical
description in the framework of the PA are omitted.
Finally, detailed calculations are performed only for the case of MNs with two layers 
(so-called duplex networks) with identical degree distributions and with full overlap of nodes
(with each node belonging to both layers) but with independently generated sets of edges; as a result, possible effects of the
correlation between the degrees of nodes within different layers and of the edge
overlap between layers \cite{Min15} on the observed phenomena are not studied.

Under the above-mentioned assumptions the model under study exhibits qualitatively similar behavior as that on MNs with layers in the form
of complete graphs \cite{Chmiel15}. However, details of the first- or second-order FM transition observed in MC
simulations (e.g., the precise location of the critical point or points, width of the possible hysteresis loop, critical exponents) 
depend significantly on the topology of the networks forming the layers of the MN. In general, better agreement between numerical and
theoretical results based on the homogeneous PA occurs for the model on MNs with layers with high density of
edges. Then, as expected, in the case of MNs composed of layers with negligible heterogeneity, e.g., RRGs and ERGs,
the details of the FM transition are quantitatively well captured by the theory based on the homogeneous PA. Besides, it is shown that
this theory yields reasonably good agreement with numerical results also for the model on MNs with 
weakly heterogeneous SF layers with finite second moment of the degree distributions. In the case of MNs with strongly
heterogeneous layers this agreement is much worse and only qualitative; possibility of formulation of heterogeneous PA
\cite{Pugliese09,Gleeson11,Gleeson13} for the model on such MNs, which could better reproduce results of MC simulations,
is only briefly discussed here. Agreement between numerical and theoretical results substantially deteriorates for the model on MNs with low
density of edges, and in this case predictions of the homogeneous PA can be quantitatively wrong even in the case of MNs with layers with negligible 
heterogeneity. Finally, it should be emphasized
that the $q$-voter model with independence is considered here only as a relatively simple example and extension of the derived PA 
to other models on MNs with similar opinion (spin) update rules is straightforward.


\section{The model}

\subsection{The $q$-voter model with independence on multiplex networks}

The $q$-voter model with independence is a sort of stochastic spin model for the opinion formation  with random sequential updating \cite{Nyczka12,Chmiel15,Jedrzejewski17}. Let us first describe the model on a monoplex network. In this model agents
represented by spins $s_{i}=\pm 1$, $i=1,2,\ldots N$ with two states corresponding to 
opposite opinions are located in $N$ nodes of the network, and the edges correspond to possible interactions between them.
The dynamics of the model is defined by the spin flip rate
which depends on a parameter $p$ ($0\le p\le 1$) determining the degree of stochasticity (``social temperature'') in the model.
This stochasticity manifests itself as agents' independence in decision making.
At each elementary time step an agent and a subset of $q$ her neighbors (a $q$-lobby) are chosen randomly;
the neighbors are chosen without repetitions. Then, the opinion of the agent is updated according to the following rule.
With probability $1-p$ the agent acts as a 
conformist and with probability $p$ acts independently. In the case of conformity the agent changes
opinion, and the spin flips, if and only if the opinions of all members of the $q$-lobby are identical and opposite to
that of the agent. In the case of independence the agent changes opinion, and the spin flips, with probability $1/2$,
independently of the opinions of the members of the $q$-lobby. Otherwise, the opinion of the agent remains unchanged. 
Hence, the elementary time step corresponds to the opinion update of one agent.
This procedure is repeated until all agents update their opinions,
which corresponds to one MCSS; thus, duration of the elementary time step is $\Delta t=1/N$.

It should be noted that the $q$-voter model with independence is a particular member of a class of noisy $q$-voter models which differ by details 
of the dynamics which is reflected in slightly different spin flip rates. For example, the agents can be allowed to act independently only if the
opinion in their neighborhood is not unanimous and otherwise they always behave as conformists \cite{Moretti13,Vieira18}, or, more generally,
the probabilities for independent action in the case of unanimous and differing opinions in the neighborhood of an agent need not be related to each other
\cite{Peralta18a,Peralta18}, or it is enough for agents to act as conformists even if only a majority fraction of their neighbors exhibits the same opinion
\cite{Vieira18}, etc. This, in turn, can lead to different critical behavior, e.g., absorbing rather than FM phase can occur even for non-zero
stochasticity level induced by independence \cite{Moretti13,Vieira18} which is not possible in the $q$-voter model with independence
as defined above. In this paper the latter model is used as a particular example, but other forms of stochasticity can be easily included in the
PA for the $q$-voter model on MNs derived below.

The $q$-voter model with independence on a MN consists of agents represented by two-state spins $s_{i}=\pm 1$,
$i=1,2,\ldots N$ located in the nodes of the MN which interact with independently chosen $q$-lobbies 
from every layer,  each $q$-lobby being a subset of $q$ agent's neighbors within one layer.
It should be emphasized that in the MN there is only one set of nodes,
while sets of edges corresponding to different layers are generated separately. Thus, the agents located in the nodes
present the same opinion $s_{i}$ to their neighbors within each layer (the agents are non-schizophrenic). The dynamics
of the model is again defined by the spin flip rate which is a generalization of that for the model on
monoplex networks. In particular, in this paper only AND
generalizations of the spin-update rule are considered in which, in general, it is assumed that a node is activated
if a sufficiently large fraction of its neighbors in every layer are active \cite{Lee14a}. Besides, different assumptions
concerning the status of independence of the agents (GLOBAL on all layers or LOCAL on each layer separately) can be 
made. The above-mentioned assumptions lead to the LOCAL\&AND or GLOBAL\&AND spin update rules which are described below. 
Of course, other generalizations of the spin flip rate are possible and tractable within the formalism of PA on MNs, e.g., 
the LOCAL\&OR spin update rule, but it seems they lead to less interesiting results \cite{Chmiel15} and will not be considered here.

In both LOCAL\&AND and GLOBAL\&AND cases the spins are updated randomly and sequentially. At each elementary time step 
an agent is chosen randomly. Then, $q$-lobbies of her neighbors, one lobby per each layer of the MN, are chosen randomly and 
independently. The neighbors
belonging to a $q$-lobby within a given layer are chosen without repetitions, however, due to the topology of connections
within the MN (see Sec.\ II.B) it can happen that the same node belongs to two or more $q$-lobbies within different layers. 

In the LOCAL\&AND case the agent first interacts with each $q$-lobby separately: with probability $1-p$ she acts as a 
conformist and tends to change opinion if and only if the opinions of all members of the $q$-lobby are identical and opposite to
her opinion, and with probability $p$ she acts independently and tends to change opinion with probability $1/2$. 
Then, if and only if for every layer of the MN the agent tends to change opinion the spin eventually flips; 
otherwise the opinion of the agent remains unchanged.
Thus, under further simplyfying assumptions used in this paper, including independence of the degree distributions within each layer (Sec. II.B),
 in the LOCAL\&AND case the spin flip rate is a product of the rates for the $q$-voter model with independence
on monoplex networks corresponding to subsequent layers of the MN.

In the GLOBAL\&AND case all $q$-lobbies from all layers of the MN are first aggregated into a single lobby. Then, 
with probability $1-p$ the agent acts as a 
conformist and changes opinion, i.e., the spin flips, if and only if the opinions of all members of the latter lobby are identical and opposite to
her opinion, and with probability $p$ the agent acts independently and changes opinion, i.e., the spin flips, with probability $1/2$. 
Otherwise, the opinion of the agent remains unchanged.
Thus, in the GLOBAL\&AND case the spin flip rate is formally equal to the rate for the $q$-voter model with independence on
an aggregate monoplex network composed of all layers of the MN with proportionally rescaled size of the $q$-lobby.  Nevertheless, there
is a small difference in the dynamics of the models on a MN and on an aggregate network: in the former case the neighborhood of a given agent 
is formed by choosing sets of $q$ neighbors, each composed of agent's neighbors within only one layer, while in the latter one the neighborhood
is formed by choosing as many neighbors as necessary without taking into account the layer in which they are connected to a given agent.
Under further simplyfying assumptions used in this paper, including independence and equality of the degree distributions within each layer (Sec. II.B),
this difference turns out to be unimportant, but in principle can manifest itself if, e.g., the degree of heterogeneity of networks forming separate layers
differs substantially.

\subsection{The models for multiplex networks}

As mentioned in Sec.\ I a MN consists of a fixed set of $N$ nodes which are connected by many separately generated sets of
edges called layers  \cite{Boccaletti14,Lee14,Lee15}. 
For simplicity, in this paper MNs with only two layers
denoted as $G^{(A)}$, $G^{(B)}$ are considered (so-called duplex networks); 
theoretical approach of Sec.\ III can be easily extended to the case with more than two layers. 
Thus, each node is characterized by two degrees $k^{(A)}$, $k^{(B)}$ defined as the numbers of edges attached to it
within the respective layer. Moreover, it is assumed that
the layers are independently generated complex networks with the degree distributions $P\left( k^{(A)}\right)$,
$P\left( k^{(B)}\right)$ and with mean degrees of nodes $\langle k^{(A)}\rangle$, $\langle k^{(B)}\rangle$.
Only MNs with layers with full overlap of nodes are considered, such that each node has non-zero degree within each layer,
$k^{(A)}>0$, $k^{(B)}>0$; thus, full overlap of nodes means that there are no nodes which belong only to one, $G^{(A)}$ or $G^{(B)}$ layer,
i.e., which have degree $k^{(B)}=0$ or $k^{(A)}=0$, respectively. Hence, the joint degree distribution is
$P\left( k^{(A)}, k^{(B)}\right) = P\left( k^{(A)}\right) P\left( k^{(B)}\right)$. In particular, this means that any correlations
exist neither between the degrees $k^{(A)}$, $k^{(B)}$ of nodes nor between the edges in the sense that the probability that two agents are neighbors
within one layer is independent of that if they are neighbors within the other layer, i.e., there is no other than accidental 
edge overlap \cite{Min15}. Consequently,
the LOCAL\&AND ar GLOBAL\&AND spin flip rates (Sec. II.A) can be, at least to some extent,
factorized into terms related to consecutive layers which facilitates
analytic calculations within the PA. Nevertheless, influence of the above-mentioned degree and edge correlations on the behavior of the
$q$-voter model on appropriately constructed MNs can be important and deserves further investigation.

Detailed theoretical calculations and MC simulations are performed for the $q$-voter model with independence
on MNs with layers characterized by identical degree distributions $P\left( k^{(A)}\right) = P\left( k^{(B)}\right)$, thus with
$\langle k^{(A)}\rangle= \langle k^{(B)}\rangle$. In particular, the layers under study can be 
complex networks in the form of RRGs, ERGs and heterogeneous SF networks. 
RRG is a sort of random graph with degree distribution $P(k)=\delta_{k,k_{0}}$ and mean degree $\langle k\rangle =k_{0}$,
with $N$ randomly connected nodes, each with the same degree $k_{0}$. 
ERG is a sort of random graph with $N$ nodes and binomial degree distribution 
$P(k)={N-1 \choose k}\tilde{p}^{k}(1-\tilde{p})^{N-1-k}$ with $\langle k\rangle=N\tilde{p}$ \cite{Albert02,Barabasi16,Erdos59}.
SF network is characterized by a power degree distribution $P(k)\propto k^{-\lambda}$, $\lambda >2$, for $k>k_{\rm min}$,
and $P(k)=0$ otherwise, with $\langle k\rangle = \left(\lambda-1\right) k_{\rm min}/\left(\lambda -2\right)$
\cite{Albert02,Barabasi16,Barabasi99}.
ERGs can be efficiently constructed by randomly connecting $N$ nodes with $N\langle k\rangle/2$ edges.
In turn, an efficient method to generate RRGs and SF networks is to apply the Configuration Model \cite{Newman03}.
Using the above-mentioned methods the two layers $G^{(A)}$, $G^{(B)}$ are generated independently. 
Care is taken to avoid multiple connections between nodes within each layer. However, since layers are generated independently
it is possible that a given pair of nodes is connected by two edges, one from the layer $G^{(A)}$ and the other from $G^{(B)}$.
Hence, in the process of MC simulation of the $q$-voter model with independence on a MN it can happen that at a certain elementary time step 
the same neighbor of a chosen agent 
belongs to both her $q$-lobbies, one within the layer $G^{(A)}$ and the other within $G^{(B)}$.


\section{Theory}

\subsection{Mean field approximation}

Before introducing homogeneous PA for the $q$-voter model on MNs with layers in the form of complex networks, in this subsection simple MFA for the
$q$-voter model on MNs with layers in the form of fully connected and overlapping graphs is recollected \cite{Chmiel15} 
and certain results which are important for further discussion are derived. In this model the neighborhood within each layer of a given spin consists of all other
spins. In the MFA the macroscopic quantity characterizing the model is the concentration $c$ of spins with direction up, 
related to the order parameter, the usual magnetization $m$, by $c=(1+m)/2$.
A dynamical equation for the concentration $c$ has a form of the rate equation,
\begin{equation}
\frac{\partial c}{\partial t}= \gamma^{+}(c,p) -\gamma^{-}(c,p),
\label{rateMF}
\end{equation}
where $\gamma^{+}$ ($\gamma^{-}$) are rates of spin flips in the direction up (down) averaged over all spins. 

In the thermodynamic limit $N\rightarrow \infty$ the rates in the case of LOCAL\&AND spin update rule are \cite{Chmiel15}
\begin{eqnarray}
\gamma^{+}(c,p)&= &(1-c) \left[ (1-p)c^{q}+\frac{p}{2}\right]^{2} \nonumber\\
\gamma^{-}(c,p)&= &c \left[ (1-p)(1-c)^{q}+\frac{p}{2}\right]^{2},
\end{eqnarray}
where in both cases the first term accounts for the fact that for the flip of the spin in the direction up (down) to occur, first a spin with direction down (up)
must be selected during a simulation, and the second term results from the application of the LOCAL\&AND spin update rule described in Sec. II.A.

It can be seen that Eq.\ (\ref{rateMF}) has a fixed point $c=1/2$ ($m=0$) corresponding to the PM phase. Denoting by 
$A(c)= (1-c) \left[ (1-p)c^{q}+\frac{p}{2}\right]^{2}- c \left[ (1-p)(1-c)^{q}+\frac{p}{2}\right]^{2}$ the right-hand side of Eq.\ (\ref{rateMF}),
with decreasing $p$ this fixed point loses stability when $\left. \frac{\partial A}{\partial c}\right|_{c=1/2} =0$, i.e.\ at 
\begin{equation}
p^{\star}_{MF}=\frac{2q-1}{2q-1+2^{q-1}}.
\label{pstarMFLA}
\end{equation}
Depending on the size of the $q$-lobby the transition from the PM to the FM phase can be first-order (for small $q$) or second-order (for large $q$). 
In the case of the second-order transition
the critical value of the independence parameter is $p_{c,MF}=p^{\star}_{MF}$, and for $p<p_{c,MF}$ Eq.\ (\ref{rateMF}) has two symmetric stable
solutions with $c<1/2$ ($m<0$) and $c>1/2$ ($m>0$), corresponding to the FM phase, and an unstable solution $c=1/2$. In the case of the 
first-order transition a hysteresis loop occurs: with decreasing $p$ the PM solution $c=1/2$ loses stability at
$p=p_{c,MF}^{(1)}=p^{\star}_{MF}$ and the only stable solutions remain the two symmetric FM ones, while for increasing $p$ the two FM solutions
disappear simultaneously at $p=p_{c,MF}^{(2)}> p_{c,MF}^{(1)}$ (the value of $p_{c,MF}^{(2)}$ can be determined numerically) and the only
stable solution is the PM one; thus, for $p_{c,MF}^{(1)}<p<p_{c,MF}^{(2)}$ there is a bistability region \cite{Chmiel15}. 

In order to gain more insight into the character of the phase transition it is convenient to rewrite Eq.\ (\ref{rateMF}) in terms of the magnetization $m$,
expand the right-hand side of the resulting rate equation in powers of $m$ and write it as a derivative of an effective potential $V(m,p,q)$,
\begin{equation}
\frac{\partial c}{\partial t}=-\frac{\partial V(m,p,q)}{\partial m},
\end{equation}
\begin{equation}
V(m,p,q)= C_{2}(p,q)m^{2}+C_{4}(p,q)m^{4}+ C_{6}(p,q)m^{6}+\ldots
\label{Vexpanded}
\end{equation}
\begin{eqnarray}
C_{2}(p,q) &=& -\frac{(1-p)^{2}}{2^{2q}}(2q-1)-\frac{p(1-p)}{2^{q}}(q-1)-\frac{p^{2}}{2}, \nonumber\\
C_{4}(p,q) &=& - \frac{(1-p)^{2}}{2^{2q+1}}\left[ {2q \choose 3} -{2q \choose 2} \right] \nonumber\\
&&- \frac{p(1-p)}{2^{q+1}}\left[ {q \choose 3} -{q \choose 2} \right],\nonumber\\
C_{6}(p,q) &=& - \frac{1}{6}\frac{(1-p)^{2}}{2^{2q-1}}\left[ {2q \choose 5} -{2q \choose 4} \right] \nonumber\\
&&- \frac{1}{6}\frac{p(1-p)}{2^{q-1}}\left[ {q \choose 5} -{q \choose 4} \right],
\end{eqnarray}
(by convention, ${x \choose y} \equiv 0$ for $x<y$ or $y<0$).
It can be easily calculated that $C_{2}(p^{\star}_{MF},q)=0$ for any $q$, $C_{4}(p^{\star}_{MF},q)>0$ for $q=2,3$, $C_{4}(p^{\star}_{MF},4)=0$
and $C_{4}(p^{\star}_{MF},q)<0$ for $q\ge 5$; moreover, $C_{6}(p^{\star}_{MF},q)>0$ for $q=4,5$. Employing formalism of the Landau theory of
phase transitions (for applications to the $q$-voter model see Refs.\ \cite{Nyczka12,Chmiel15}) 
it can be deduced that second-order transition occurs in the model under
study at $p=p_{c,MF}=p^{\star}_{MF}$ for $q=2,3$. On the other hand, for $q\ge 5$ first-order transition is expected provided that 
at $p=p^{\star}_{MF}$ a pair of symmetric minima with $\left| m\right|>0$ of the potential $V(m,p,q)$ exists, corresponding to the 
coexisting stable FM phase; for $q=5$ this is guaranteed by
$C_{6}(p^{\star}_{MF},5)>0$ while for $q\ge 6$ higher-order nonlinear terms in the expansion (\ref{Vexpanded}) must be responsible for this.
At $p=p^{\star}_{MF}$ and $q=4$, due to mutual disappearance of the coefficients $C_{2}$ and $C_{4}$ with $C_{6}>0$,
a tricritical point is expected separating the first- and second-order phase transition curves on the $(p,q)$ plane. 
The above results 
concerning the critical values $p_{c,MF}$ or $p_{c,MF}^{(1)}$ as well as the order of the FM transition are in agreement with numerical analysis
of the MF equation (\ref{rateMF}) and direct MC simulation of the $q$-voter model with LOCAL\&AND spin update rue 
on a MN with layers in the form of complete graphs \cite{Chmiel15}. 
In particular, comparison of Eq.\ (\ref{pstarMFLA}) for $q=2$ ($p_{MF}^{\star}=0.6$) and $q=3$ ($p_{MF}^{\star}=5/9=0.555\ldots$) with Fig.\ 9 and Fig.\
16(b) of Ref.\ \cite{Chmiel15} (when the model shows continuous phase transition) and for $q=5$ ($p_{MF}^{\star}=0.36$) with Fig.\ 14 and Fig.\
16(b) of Ref.\ \cite{Chmiel15} (when the model shows first-order phase transition) shows good agreement of Eq.\ (\ref{pstarMFLA}) with predictions
based on the MFA and with results of MC simulations of the respective $q$-voter model with independence.

In the case of the GLOBAL\&AND spin update rule the rates in Eq.\ (\ref{rateMF}) in the thermodynamic limit are \cite{Chmiel15}
\begin{eqnarray}
\gamma^{+}(c,p)&= &(1-c) \left[ (1-p)c^{2q}+\frac{p}{2}\right] \nonumber\\
\gamma^{-}(c,p)&= &c \left[ (1-p)(1-c)^{2q}+\frac{p}{2}\right],
\end{eqnarray}
thus in the MFA the $q$-voter model on a MN with fully connected layers is equivalent to the $2q$-voter model on a (monoplex)
fully connected graph. The fixed point with $c=1/2$ loses stability at
\begin{equation}
p_{MF}^{\star}= \frac{2q-1}{2q-1+2^{2q-1}},
\label{pstarMFGA}
\end{equation}
which may be also obtained from the result for the $q$-voter model on a fully connected graph, Eq.\ (26) in Ref.\ \cite{Nyczka12}, simply by changing
$q$ into $2q$. 
Comparison of the above result for $q=2$ ($p_{MF}^{\star}=0.2727\ldots$) with Fig.\ 7 of Ref.\ \cite{Chmiel15}
(when the model shows continuous phase transition) shows good agreement of Eq.\ (\ref{pstarMFGA}) with numerical predictions
based on the MFA and with results of MC simulations. The coefficients in the expansion of the effective potential (\ref{Vexpanded}) are
\begin{eqnarray}
C_{2}(p,q) &=&- \frac{1-p}{2^{2q}}(2q-1)-p, \nonumber\\
C_{4}(p,q) &=&  -\frac{1-p}{2^{2q+1}}\left[ {2q \choose 3} -{2q \choose 2} \right] \\
C_{6}(p,q) &=& -\frac{1}{6} \frac{1-p}{2^{2q-1}}\left[ {2q \choose 5} -{2q \choose 4} \right].
\end{eqnarray}
As expected, $C_{2}(p^{\star}_{MF},q)=0$ for any $q$, $C_{4}(p^{\star}_{MF},q)>0$ for $q<5/2$ and $C_{4}(p^{\star}_{MF},q)<0$ for $q>5/2$,
thus for $q=2$ the second-order FM transition occurs at $p=p_{c,MF}=p^{\star}_{MF}$, and the transition should be first-order for $q\ge 3$. 
Since the size of the $q$-lobby is an integer number there is no tricritical point separating the first- and second-order FM transition curves on the
$(p,q)$ plane characterized by mutual disappearance of the $C_{2}$ and $C_{4}$ coefficients.

Predictions of the above simple MFA concerning the order of the FM transition for different $q$ and the existence of the
tricritical point turn out to be qualitatively correct also in the case of the $q$-voter model on MNs with layers in the form of complex networks. Besides,
quantitative predictions of the more advanced homogeneous PA formulated below also converge to those of the MFA in the limit 
$\langle k^{(A)}\rangle, \langle k^{(B)}\rangle \rightarrow \infty$, as expected.

\subsection{General formulation of the pair approximation}

Theoretical description of the $q$-voter model with independence and $q$-neighbor Ising model on MNs 
with layers in the form of complete graphs based on the MFA
yields quantitative agreement with results of MC simulations concerning both the order of the transition from the PM to the FM phase and
the position of the critical point \cite{Chmiel15,Chmiel17}. However, when the above-mentioned models are studied on monoplex
networks a more accurate PA is necessary in order to take into account the effect of the network topology (e.g., the degree
distribution and the mean degree of nodes) on the properties of the phase transition \cite{Jedrzejewski17,Peralta18,Chmiel18}. 
Hence, in this paper PA is used for theoretical investigation of the $q$-voter model with independence on MNs with two layers
in the form of complex networks.
In order to make our approach more widely applicable in this subsection a general case of a two-state spin model with up-down symmetry 
is considered, and application of the results to the $q$-voter model with 
LOCAL\&AND or GLOBAL\&AND spin update rules is presented in Sec.\ III.B.
Henceforth in this subsection $\hat{p}$ denotes a parameter which controls the degree of stochasticity in the general model
(in the case of the $q$-voter model this is the parameter $p$ characterizing the degree of independence of agents).

In the framework of the PA macroscopic quantities characterizing the model are
 concentrations of spins with orientation up or down
located in nodes with degrees $k^{(A)}$, $k^{(B)}$, denoted as
$c_{k^{(A)},k^{(B)},j} = P\left(\left. j\right| k^{(A)},k^{(B)}\right)$, $j \in \left\{ \uparrow, \downarrow \right\}$ (hence,
$c_{k^{(A)},k^{(B)},\downarrow} = 1- c_{k^{(A)},k^{(B)},\uparrow}$), as well as concentrations (within separate layers of the MN)
of active bonds connecting nodes occupied by spins 
with opposite orientations. In the heterogeneous PA it should be taken into account that the latter concentrations depend on the degrees of the connected 
nodes \cite{Pugliese09,Gleeson11,Gleeson13}. As mentioned in Sec.\ I, in this paper only homogeneous PA is considered in which only average concentrations of active bonds
$b^{(A)}$, $b^{(B)}$ within the layers $G^{(A)}$, $G^{(B)}$
normalized to the total numbers of edges $N\langle k^{(A)}\rangle/2$, $N\langle k^{(B)}\rangle/2$, respectively,
independent of the degrees of connected nodes, are assumed as macroscopic quantities. Due to this assumption possible effect of heterogeneity 
of the layers on the properties of the phase transitions observed in the model is neglected to large extent, 
which can be particularly strong in the case of layers in the form of SF networks with $2< \lambda \le 3$. However, for the layers in the form of
weakly heterogeneous RRGs, ERGs and SF networks with $\lambda>3$ the homogeneous PA is expected to yield quantitatively correct results. Moreover,
in the homogeneous PA the number of equations for the significant macroscopic quantities is radically reduced, which in the case of the $q$-voter model 
with independence enables one to predict, partly analytically, the location of the critical point and the stability of the PM and FM phases, similarly as in the MFA. 

In the framework of the homogeneous PA the probabilities that a randomly selected spin with orientation $j \in \left\{ \uparrow, \downarrow \right\}$
has within the layer $G^{(A)}$ ($G^{(B)}$) a neighbor with opposite orientation are independent of the degrees of nodes and
are dentoted by $\theta_{j}^{(A)}$ ($\theta_{j}^{(B)}$). These probabilities play a crucial role in the formulation of the dynamical equations of the PA
and can be expressed by the above-mentioned macroscopic concentrations by generalizing reasoning for monoplex networks \cite{Peralta18a}  in the
following way. In the case of  layers with full overlap of nodes the number of attachments of active bonds to nodes with spins with a given orientation $j$,
independently of their degrees, within a given layer, say $G^{(A)}$, is $N\langle k^{(A)} \rangle b^{(A)}/2$ and the number of attachments of all bonds
to such nodes is $\sum_{k^{(A)},k^{(B)}} NP\left( k^{(A)},k^{(B)}\right)k^{(A)} c_{k^{(A)},k^{(B)},j}$.
Indroducing the ``weighted'' or ``link'' layer-dependent concentration of nodes with spins up 
$C_{\uparrow}^{(A)} = \sum_{k^{(A)},k^{(B)}} P\left( k^{(A)},k^{(B)}\right)k^{(A)} c_{k^{(A)},k^{(B)},\uparrow}/\langle k^{(A)}\rangle$, 
taking into account that
the respective concentration of nodes with spin down is $C_{\downarrow}^{(A)}=1-C_{\uparrow}^{(A)}$, and repeating the above reasoning 
for the layer $G^{(B)}$ with the weighted concentration of nodes with spins up $C_{\uparrow}^{(B)}$ the conditional probabilities
$\theta_{j}^{(A)}$,  $\theta_{j}^{(B)}$, $j \in \left\{ \uparrow, \downarrow \right\}$,
are finally obtained as 
\begin{eqnarray}
    \theta_{\uparrow}^{(A)}=\frac{b^{(A)}}{2C_{\uparrow}^{(A)}}, \qquad \theta_{\downarrow}^{(A)}=\frac{b^{(A)}}{2\left(1-C_{\uparrow}^{(A)}\right)}, \nonumber \\
    \theta_{\uparrow}^{(B)}=\frac{b^{(B)}}{2C_{\uparrow}^{(B)}}, \qquad \theta_{\downarrow}^{(B)}=\frac{b^{(B)}}{2\left(1-C_{\uparrow}^{(B)}\right)}.   
\label{condprobspins}
\end{eqnarray}  

In the case of models on MNs the dynamical equations of the homogeneous PA comprise equations for the concentrations of nodes with spins up
$c_{k^{(A)},k^{(B)},\uparrow}$ as well as for concentrations of active links within the layers $b^{(A)}$, $b^{(B)}$. 
The former equations can be written in a general form as rate equations,
\begin{eqnarray}
\frac{\partial c_{k^{(A)},k^{(B)},\uparrow}}{\partial t} &=&
\gamma^{+} \left( c_{k^{(A)},k^{(B)},\uparrow},\theta_{\downarrow}^{(A)},\theta_{\downarrow}^{(B)}, \hat{p}\right) \nonumber\\
&-&\gamma^{-} \left( c_{k^{(A)},k^{(B)},\uparrow},\theta_{\uparrow}^{(A)},\theta_{\uparrow}^{(B)}, \hat{p}\right), 
\label{rate1}
\end{eqnarray}
where $\gamma^{+}$ ($\gamma^{-}$) are rates of spin flips in the direction up (down) averaged over all nodes with degrees $k^{(A)}$, $k^{(B)}$.
In many cases, including the $q$-voter model on MNs considered in this paper, the latter rates can be decomposed into products 
\begin{eqnarray}
\gamma^{+}  &=&\left( 1- c_{k^{(A)},k^{(B)},\uparrow}\right) 
F^{+} \left(\left. \theta_{\downarrow}^{(A)},\theta_{\downarrow}^{(B)}, \hat{p}\right| \downarrow\right),
\nonumber\\
\gamma^{-}  &=& c_{k^{(A)},k^{(B)},\uparrow} F^{-} \left(\left. \theta_{\uparrow}^{(A)},\theta_{\uparrow}^{(B)}, \hat{p}\right|\uparrow\right).
\label{ratesproduct}
\end{eqnarray}
In Eq.\ (\ref{ratesproduct}) 
the concentrations of nodes with spin up (down) account for the fact that in order to increase (decrease) the concentration of nodes with spins up
during a simulation first a spin with orientation down (up) must be selected and its flip should be attempted. The functions $F^{+}$ ($F^{-}$) are 
rates of spin flips provided that a spin with orientation down (up) was selected, which depend on the model under study.
Note that due to the homogeneous PA, and the particular model under study, the latter functions do not depend
on the degrees $k^{(A)}$, $k^{(B)}$. Then, multiplying both sides of Eq.\ (\ref{rate1}) by 
$P\left( k^{(A)},k^{(B)}\right) k^{(A)}/\langle k^{(A)}\rangle$ and summing over $k^{(A)}$ it is possible to obtain a dynamical equation for the weighted
concentration $C_{\uparrow}^{(A)}$ (an equation for $C_{\uparrow}^{(B)}$ is derived analogously),
\begin{eqnarray}
\frac{\partial C_{\uparrow}^{(A)}}{\partial t} &=&
\left(1-C_{\uparrow}^{(A)}\right) 
F^{+} \left(\left. \theta_{\downarrow}^{(A)},\theta_{\downarrow}^{(B)}, \hat{p}\right| \downarrow\right) \nonumber\\
&-&C_{\uparrow}^{(A)} F^{-} \left(\left. \theta_{\uparrow}^{(A)},\theta_{\uparrow}^{(B)}, \hat{p}\right|\uparrow\right). 
\label{rate1A}
\end{eqnarray}

Moreover, it can be easily shown from Eq.\ (\ref{rate1}, \ref{ratesproduct}) 
that for any different degrees $c_{k^{(A)},k^{(B)},\uparrow} -  c_{k^{\prime (A)},k^{\prime (B)},\uparrow} \rightarrow 0$ with increasing time,
thus the systems of equations (\ref{rate1}) and (\ref{rate1A}) both have a stable stationary solution
$c_{k^{(A)},k^{(B)},\uparrow}=c$ for any degrees $k^{(A)},k^{(B)}$ as well as $C_{\uparrow}^{(A)}=C_{\uparrow}^{(B)}=c$ , where 
 $c$ is an usual (unweighted) concentration of nodes with spins up in the MN; 
since the nodes of layers of the MN fully overlap a single quantity $c$ yields the concentrations of nodes with spins up within each layer,
as expected. As mentioned in Sec.\ I in this paper we are interested only in the stationary solutions (fixed points) of the system of equations for the macroscopic
quantities, characterizing different thermodynamic phases in the model. Hence, in the adiabatic limit for long times the concentrations
$c_{k^{(A)},k^{(B)},\uparrow} $ can be replaced with $c$ \cite{Peralta18a} and Eq.\ (\ref{rate1A}) can be written in the form
\begin{eqnarray}
\frac{\partial c}{\partial t} &=&
(1-c) F^{+} \left(\left. \theta_{\downarrow}^{(A)},\theta_{\downarrow}^{(B)}, \hat{p}\right| \downarrow\right) \nonumber\\
&-& c F^{-} \left(\left. \theta_{\downarrow}^{(A)},\theta_{\downarrow}^{(B)}, \hat{p}\right|\uparrow\right). 
\label{rate1B}
\end{eqnarray}
In the same approximation the conditional probabilities in Eq.\ (\ref{condprobspins}) become
\begin{eqnarray}
    \theta_{\uparrow}^{(A)}=\frac{b^{(A)}}{2c}, \qquad \theta_{\downarrow}^{(A)}=\frac{b^{(A)}}{2(1-c)}, \nonumber \\
    \theta_{\uparrow}^{(B)}=\frac{b^{(B)}}{2c}, \qquad \theta_{\downarrow}^{(B)}=\frac{b^{(B)}}{2(1-c)},
\label{condprobspinsA}   
\end{eqnarray}    
and resemble those appearing in the PA for the $q$-voter model on monoplex networks \cite{Jedrzejewski17}.

The dynamical equations for the concentrations of active links can be obtained by observing that
each flip of a spin located in the node with degrees $k^{(A)}$, $k^{(B)}$ causes changes in $b^{(A)}$ and $b^{(B)}$ by 
\begin{eqnarray}
\Delta_{b^{(A)}} \left( i^{(A)}\left| k^{(A)}\right.\right)&=&\frac{2}{N\langle k^{(A)}\rangle}\left( k^{(A)}-2i^{(A)} \right), \nonumber\\
\Delta_{b^{(B)}} \left( i^{(B)}\left| k^{(B)}\right.\right)&=&\frac{2}{N\langle k^{(B)}\rangle}\left( k^{(B)}-2i^{(B)} \right),
\end{eqnarray}
where $i^{(A)}$ ($i^{(B)}$) is the number of active bonds attached to this node within the layer $G^{(A)}$ ($G^{(B)}$), 
$0\le i^{(A)}\le k^{(A)}$, $0\le i^{(B)}\le k^{(B)}$. The spin flip rate 
$f\left( i^{(A)},i^{(B)},\hat{p}\left| k^{(A)},k^{(B)}\right.\right)$ is also
a function of the concentrations of active links within layers, and its form characterizes the particular model. Denoting by 
$P\left( j, i^{(A)},i^{(B)}\left| k^{(A)},k^{(B)}\right.\right)$ conditional probability that the spin with 
orientation $j \in \left\{ \uparrow, \downarrow \right\}$  
has $i^{(A)}$, $i^{(B)}$ neighboring spins  with opposite orientation within the layers $G^{(A)}$, $G^{(B)}$, respectively, 
provided that it is located in the node with degrees $k^{(A)}$, $k^{(B)}$, it is easily
obtained that the average changes of the concentrations $b^{(A)}$, $b^{(B)}$ at each elementary time step are
\begin{widetext}
    \begin{eqnarray}
        \Delta b^{(A)}&=&  \sum_{j\in \left\{ \uparrow,\downarrow\right\}}
        \sum_{k^{(A)},k^{(B)}} P\left( k^{(A)}, k^{(B)}\right) 
        \sum_{i^{(A)}=0}^{k^{(A)}} \sum_{i^{(B)}=0}^{k^{(B)}} 
        P\left( j, i^{(A)},i^{(B)}\left| k^{(A)},k^{(B)}\right.\right)
        \nonumber\\
        &\times&
        f\left( i^{(A)},i^{(B)},\hat{p}\left| k^{(A)},k^{(B)}\right.\right) 
        \Delta_{b^{(A)}} \left( i^{(A)}\left| k^{(A)}\right.\right),
        \label{DeltabA}
    \end{eqnarray}
\end{widetext}
and $\Delta b^{(B)}$ which is evaluated as in Eq.\ (\ref{DeltabA}) with
$\Delta_{b^{(A)}} \left( i^{(A)}\left| k^{(A)}\right.\right)$ replaced by $\Delta_{b^{(B)}} \left( i^{(B)}\left| k^{(B)}\right.\right)$.
In the first approximation it can be assumed that orientations of the above-mentioned neighboring spins are independent.
In this approximation the possibility is neglected that a pair of neighboring spins can be also connected by an edge in one or both layers
$G^{(A)}, G^{(B)}$ and thus interact directly. Moreover, the possibility is neglected that a neighboring spin in the layer $G^{(A)}$
is also a neighboring spin in the layer $G^{(B)}$ if the respective edges within these two layers overlap and form a multiple connection
within the MN (Sec.\ II.B). 
As a result it can be approximated that the numbers of active bonds $i^{(A)}$, $i^{(B)}$
attached to the node with degrees $k^{(A)}$, $k^{(B)}$ within the layers $G^{(A)}$, $G^{(B)}$ 
result from independent binomial distributions with parameters given by Eq.\ (\ref{condprobspinsA}), thus
    \begin{eqnarray}
&&P\left( j, i^{(A)},i^{(B)}\left| k^{(A)},k^{(B)}\right.\right) \nonumber\\ 
&=& P\left(\left. j\right| k^{(A)},k^{(B)}\right)  P\left( i^{(A)},i^{(B)}\left| k^{(A)},k^{(B)},j\right.\right) \nonumber\\
&=& c_{k^{(A)},k^{(B)},j} B_{k^{(A)},i^{(A)}}\left( \theta_{j}^{(A)}\right) B_{k^{(B)},i^{(B)}}\left( \theta_{j}^{(B)}\right),
        \label{Pbinomial}
    \end{eqnarray}
where $B_{k,i}( \theta) = {k \choose i} \theta^{i} (1-\theta)^{k-i}$ denotes the binomial factor. 
Again approximating $ c_{k^{(A)},k^{(B)},\uparrow} \approx c$,
combining Eq.\ (\ref{DeltabA},\ref{Pbinomial}), 
taking into account that an elementary time step is $\Delta t =1/N$ and going to
the thermodynamic limit $N\rightarrow \infty$ yields 
the equation for the concentration $b^{(A)}$
\begin{widetext}
\begin{eqnarray}
\frac{\partial b^{(A)}}{\partial t} &=& \frac{2}{\langle k^{(A)}\rangle} \sum_{j\in \left\{ \uparrow,\downarrow\right\}} c_{j}
\sum_{k^{(A)},k^{(B)}} P\left( k^{(A)}, k^{(B)}\right) 
\sum_{i^{(A)}=0}^{k^{(A)}} \sum_{i^{(B)}=0}^{k^{(B)}} 
 B_{k^{(A)},i^{(A)}}\left( \theta_{j}^{(A)}\right) B_{k^{(B)},i^{(B)}}\left( \theta_{j}^{(B)}\right)
\nonumber\\
&\times& f\left( i^{(A)},i^{(B)},\hat{p}\left| k^{(A)},k^{(B)}\right.\right) \left( k^{(A)}-2i^{(A)}\right), \label{rate2}
\end{eqnarray}
\end{widetext}
and a complementary equation for the concentration $b^{(B)}$ which can be obtained from Eq.\ (\ref{rate2}) by replacing
$\langle k^{(A)}\rangle$ by $\langle k^{(B)}\rangle$ and $ k^{(A)}-2i^{(A)}$ by $ k^{(B)}-2i^{(B)}$.

As mentioned above, homogeneous PA under certain quite general assumptions leads to a radical decrease of the number of dynamical equations for the significant
macroscopic quantities; in the case of models on MNs with full overlap of nodes
there are only three such quantities, i.e., the concentration $c$ of nodes with spins up
in the MN and concentrations $b^{(A)}$, $b^{(B)}$ of active links within the consecutive layers $G^{(A)}$, $G^{(B)}$. Further reduction is possible under certain 
assumptions concerning the degree distributions of the layers. 

A final form of the system of equations (\ref{rate2}) depends on the joint degree distribution 
$P\left( k^{(A)}, k^{(B)}\right) $ and the form of the spin flip rate.
A particularly simple situation occurs if the layers 
of the MN are independently generated networks
with degree distributions $P\left( k^{(A)}\right)$, $P\left( k^{(B)}\right)$ which yields the joint degree distribution
$P\left( k^{(A)}, k^{(B)}\right)= P\left( k^{(A)}\right) P\left( k^{(B)}\right)$, and if the model obeys the LOCAL\&AND
spin update rule so that the spin flip rate is a product of rates for the model on two monoplex networks corresponding to
the two layers, 
$f\left( i^{(A)},i^{(B)},\hat{p}\left| k^{(A)},k^{(B)}\right.\right)=$
$f\left( i^{(A)},\hat{p}\left| k^{(A)}\right.\right) f\left( i^{(B)},\hat{p}\left| k^{(B)}\right.\right)$.
Under these two assumptions 
the summations over $k^{(A)}$, $k^{(B)}$ as well as over
$i^{(A)}$, $i^{(B)}$ in Eq.\ (\ref{rate2}) can be performed separately which yields
\begin{widetext}
    \begin{eqnarray}
 \frac{\partial b^{(A)}}{\partial t} &=& \frac{2}{\langle k^{(A)}\rangle} \sum_{j\in \left\{ \uparrow,\downarrow\right\}} c_{j}       
        \sum_{k^{(A)}} P\left( k^{(A)}\right)  
        \sum_{i^{(A)}=0}^{k^{(A)}}
        B_{k^{(A)},i^{(A)}}\left( \theta_{j}^{(A)}\right) 
        f\left( i^{(A)},\hat{p}\left| k^{(A)}\right.\right) \left( k^{(A)}-2i^{(A)}\right) \nonumber\\
        &\times& 
        \sum_{k^{(B)}} P\left( k^{(B)}\right)  
        \sum_{i^{(B)}=0}^{k^{(B)}}
        B_{k^{(B)},i^{(B)}}\left( \theta_{j}^{(B)}\right)
        f\left( i^{(B)},\hat{p}\left| k^{(B)}\right.\right),
        \label{rate3}
    \end{eqnarray}
\end{widetext}
    and a complementary equation for $b^{(B)}$ which can be obtained from Eq.\ (\ref{rate3}) by exchanging the
    superscripts $A$, $B$.

\subsection{Application to the case of the $q$-voter model with independence on multiplex networks}

In this subsection general equations of Sec.\ III.A for the concentrations of spins with orientation up and of the active bonds are
written explicitly for the $q$-voter model with independence on MNs with two layers. 

Let us start with the model with the LOCAL\&AND spin update rule.
For the $q$-voter model with independence on a monoplex network corresponding to
the layer $G^{(A)}$ of the MN the spin flip rate is
\begin{eqnarray}
f\left( i^{(A)},p\left| k^{(A)}\right.\right) &=&
(1-p) \frac{\prod_{j=1}^{q}\left( i^{(A)}-j+1\right)}{\prod_{j=1}^{q}\left( k^{(A)}-j+1\right)} +\frac{p}{2}
\nonumber\\
&=&
(1-p)\frac{i^{(A)}!\left( k^{(A)} -q\right)!}{k^{(A)}!\left(i^{(A)}-q\right)!} +\frac{p}{2}.
\label{fkv}
\end{eqnarray}
The formula for $f\left( i^{(B)},\hat{p}\left| k^{(B)}\right.\right)$ 
can be obtained from Eq.\ (\ref{fkv}) by changing the superscript $A$ into $B$. 
The averaged rates $\gamma^{+}$, $\gamma^{-}$ in Eq.\ (\ref{rate1}) 
for nodes with given degrees $k^{(A)}$, $k^{(B)}$ can be obtained as averages of the spin flip rate, Eq.\ (\ref{fkv}), over the binomial distributions of the 
numbers of active bonds attached to such nodes within the layers $G^{(A)}$ and $G^{(B)}$. Since 
$\sum_{i^{(A)}=0}^{k^{(A)}} B_{k^{(A)},i^{(A)}}\left( \theta_{j}^{(A)}\right) \frac{i^{(A)}!\left( k^{(A)} -q\right)!}{k^{(A)}!\left(i^{(A)}-q\right)!}= \theta_{j}^{q}$, etc. \cite{Jedrzejewski17}, 
the rates assume the form as in Eq.\ (\ref{ratesproduct}),
    \begin{eqnarray}
        && \gamma^{+} \left( c_{k^{(A)},k^{(B)},\uparrow},\theta_{\downarrow}^{(A)},\theta_{\downarrow}^{(B)}, p\right)
        \nonumber\\
       && = \left( 1- c_{k^{(A)},k^{(B)},\uparrow}\right) \left[ (1-p)\theta_{\downarrow}^{(A)q} +\frac{p}{2}\right]
        \left[ (1-p)\theta_{\downarrow}^{(B)q} +\frac{p}{2}\right], 
        \nonumber \\
        && \gamma^{-} \left( c_{k^{(A)},k^{(B)},\uparrow},\theta_{\uparrow}^{(A)},\theta_{\uparrow}^{(B)}, p\right)
        \nonumber\\
        && = c_{k^{(A)},k^{(B)},\uparrow} \left[ (1-p)\theta_{\uparrow}^{(A)q} +\frac{p}{2}\right]
        \left[ (1-p)\theta_{\uparrow}^{(B)q} +\frac{p}{2}\right].
        \label{gammapm}
    \end{eqnarray}

Substituting Eq.\ (\ref{gammapm}) in Eq.\ (\ref{rate1}) and using Eq.\ (\ref{rate1B}) as well as substituting
Eq.\ (\ref{fkv}) in Eq.\ (\ref{rate3}) and performing summations over $k^{(A)}$, $k^{(B)}$ as in Ref.\ \cite{Jedrzejewski17} 
yields the following system of 
equations for the macroscopic quantities $c$, $b^{(A)}$, $b^{(B)}$ with $\gamma^{\pm}$ given by Eq.\ (\ref{gammapm})
with $c_{k^{(A)},k^{(B)},\uparrow}$ replaced by $c$,
\begin{widetext}
    \begin{eqnarray}
        \frac{\partial c}{\partial t} &=&\gamma^{+} \left( c,\theta_{\downarrow}^{(A)},\theta_{\downarrow}^{(B)}, p\right)
        -\gamma^{-} \left( c,\theta_{\uparrow}^{(A)},\theta_{\uparrow}^{(B)}, p\right), 
        \label{systemcb3v} \nonumber\\
        \frac{\partial b^{(A)}}{\partial t} &=& \frac{2}{\langle k^{(A)}\rangle} \sum_{j\in \left\{ \uparrow,\downarrow\right\}} c_{j} 
        \left\{ (1-p) \theta_{j}^{(A)q}
        \left[  \langle k^{(A)}\rangle-2q -2 \left( \langle k^{(A)}\rangle -q\right) \theta_{j}^{(A)} \right]
        +\frac{p}{2}\langle k^{(A)}\rangle \left(1-2 \theta_{j}^{(A)}\right) \right\} 
        \left[ (1-p) \theta_{j}^{(B)q} +\frac{p}{2}\right], \nonumber\\
        \label{systemcb4v}
    \end{eqnarray}
\end{widetext}
    and a complementary equation for $b^{(B)}$ obtained from Eq.\ (\ref{systemcb4v}) by exchanging the 
    superscripts $A$ and $B$.

Further simplification of the above system of equations
can be achieved if the two layers are independently generated
networks with identical degree distributions $P\left( k^{(A)}\right) =P\left( k^{(B)}\right)$ 
and mean degrees of nodes
$\langle k^{(A)} \rangle = \langle k^{(B)}\rangle = \langle k\rangle$. Due to the symmetry of Eq.\ (\ref{systemcb4v}) and
the complementary equation for $b^{(B)}$ a solution exists with equal active bond concentrations within both layers,
$b^{(A)}=b^{(B)} =b$ (and thus with $\theta_{j}^{(A)} =\theta_{j}^{(B)} =\theta_{j}$, 
$j\in \left\{ \uparrow,\downarrow\right\}$). 
Then the two rate equations for the active bond concentrations can be replaced with a single equation for $b$.
Hence, in this case the model on a MN is described by only two macroscopic 
quantities $c$, $b$ obeying the equations
\begin{widetext}
    
    \begin{eqnarray}
        \frac{\partial c}{\partial t} &=&(1-c) \left[ (1-p)\theta_{\downarrow}^{q} +\frac{p}{2}\right]^2 -
        c\left[ (1-p)\theta_{\uparrow}^{q} +\frac{p}{2}\right]^2, \nonumber \\
        \frac{\partial b}{\partial t} &=& \frac{2}{\langle k\rangle} \sum_{j\in \left\{ \uparrow,\downarrow\right\}} c_{j}
        \left\{ (1-p) \theta_{j}^{q}
        \left[  \langle k\rangle-2q -2 \left( \langle k\rangle -q\right) \theta_{j} \right]
        +\frac{p}{2}\langle k\rangle \left(1-2 \theta_{j}\right) \right\}
        \left[ (1-p) \theta_{j}^{q} +\frac{p}{2}\right].
        \label{systemcb5v}
    \end{eqnarray}
\end{widetext}
    In particular, for $p=0$ the above system of equations reduces to
    \begin{eqnarray}
        \frac{\partial c}{\partial t} &=&(1-c) \theta_{\downarrow}^{2q} - c \theta_{\uparrow}^{2q}, \nonumber \\
\frac{\partial b}{\partial t} &=& \frac{2}{\langle k\rangle} \sum_{j\in \left\{ \uparrow,\downarrow\right\}} c_{j} 
\theta_{j}^{2q}
\left[  \langle k\rangle-2q -2 \left( \langle k\rangle -q\right) \theta_{j} \right].
\label{systemcb6v}
\end{eqnarray}
It can be easily verified that 
in the framework of the PA the above $q$-voter model with $p=0$ is equivalent to 
the $2q$-voter model on an aggregate monoplex network 
with mean degree $2\langle k\rangle$ 
being a superposition of the two layers. However, this is not the case for $0<p \le 1$.

In the case of the GLOBAL\&AND spin update rule the respective
spin flip rate cannot be written as a product of the rates evaluated separately for each layer. 
For the model on MNs with two layers it takes a form
\begin{widetext}
    \begin{eqnarray}
        f\left( i^{(A)},i^{(B)},\hat{p}\left| k^{(A)},k^{(B)}\right.\right) &=&
        (1-p) \frac{\prod_{j=1}^{q}\left( i^{(A)}-j+1\right)}{\prod_{j=1}^{q}\left( k^{(A)}-j+1\right)} 
        \frac{\prod_{j^{\prime}=1}^{q}\left( i^{(B)}-j^{\prime}+1\right)}{\prod_{j^{\prime}=1}^{q}\left( k^{(B)}-j^{\prime}+1\right)} 
        +\frac{p}{2}
        \nonumber\\
        &=&
        (1-p)\frac{i^{(A)}!\left( k^{(A)} -q\right)!}{k^{(A)}!\left(i^{(A)}-q\right)!} 
        \frac{i^{(B)}!\left( k^{(B)} -q\right)!}{k^{(B)}!\left(i^{(B)}-q\right)!}
        +\frac{p}{2}.
        \label{fkvG}
    \end{eqnarray}
\end{widetext}
The rates
$\gamma^{+}$, $\gamma^{-}$  in Eq.\ (\ref{ratesproduct}) are
\begin{eqnarray}
\gamma^{+} \left( c_{k^{(A)},k^{(B)},\uparrow},\theta_{\downarrow}^{(A)},\theta_{\downarrow}^{(B)}, p\right)
&=& \left(1-c_{k^{(A)},k^{(B)},\uparrow}\right) \left[ (1-p)\theta_{\downarrow}^{(A)q}\theta_{\downarrow}^{(B)q} +\frac{p}{2}\right], 
\nonumber \\
\gamma^{-} \left( c_{k^{(A)},k^{(B)},\uparrow},\theta_{\uparrow}^{(A)},\theta_{\uparrow}^{(B)}, p\right)
&=& c_{k^{(A)},k^{(B)},\uparrow} \left[ (1-p)\theta_{\uparrow}^{(A)q} \theta_{\uparrow}^{(B)q}+\frac{p}{2}\right].
\label{gammaplusvG}
\end{eqnarray}
Assuming again that layers of the MN are independently generated so that 
$P\left( k^{(A)}, k^{(B)}\right)= P\left( k^{(A)}\right) P\left( k^{(B)}\right)$, 
substituting Eq.\ (\ref{gammaplusvG}) in Eq.\ (\ref{rate1}) and using Eq.\ (\ref{rate1B})
as well as substituting Eq.\ (\ref{fkvG}) in Eq.\ (\ref{rate2}) and in the latter case performing summations 
over $k^{(A)}$, $k^{(B)}$ as in Ref.\ \cite{Jedrzejewski17} 
yields the following system of
equations for the macroscopic quantities $c$, $b^{(A)}$, $b^{(B)}$, 
with $\gamma^{\pm}$ given by Eq.\ (\ref{gammaplusvG})
with $c_{k^{(A)},k^{(B)},\uparrow}$ replaced by $c$,
\begin{widetext}
    
    \begin{eqnarray}
        \frac{\partial c}{\partial t} &=&\gamma^{+} \left( c,\theta_{\downarrow}^{(A)},\theta_{\downarrow}^{(B)}, p\right)
        -\gamma^{-} \left( c,\theta_{\uparrow}^{(A)},\theta_{\uparrow}^{(B)}, p\right), \nonumber
        \label{systemcb3vG}\\
        \frac{\partial b^{(A)}}{\partial t} &=& \frac{2}{\langle k^{(A)}\rangle} \sum_{j\in \left\{ \uparrow,\downarrow\right\}} c_{j} 
        \left\{ (1-p) \theta_{j}^{(A)q} \theta_{j}^{(B)q}
        \left[  \langle k^{(A)}\rangle-2q -2 \left( \langle k^{(A)}\rangle -q\right) \theta_{j}^{(A)} \right]
        +\frac{p}{2}\langle k^{(A)}\rangle \left(1-2 \theta_{j}^{(A)}\right) \right\},
        \label{systemcb4vG}
    \end{eqnarray}
\end{widetext}
and a complementary equation for $b^{(B)}$ obtained from Eq.\ (\ref{systemcb4vG}) by exchanging the 
superscripts $A$ and $B$.

In the case of independently generated layers with identical degree distributions $P\left( k^{(A)}\right) =P\left( k^{(B)}\right)$
and $\langle k^{(A)} \rangle = \langle k^{(B)}\rangle = \langle k\rangle$ the solution with $b^{(A)}=b^{(B)} =b$ exists which obeys a system of equations
\begin{widetext}
    
    \begin{eqnarray}
        \frac{\partial c}{\partial t} &=&(1-c) \left[ (1-p)\theta_{\downarrow}^{2q} +\frac{p}{2}\right] -
        c\left[ (1-p)\theta_{\uparrow}^{2q} +\frac{p}{2}\right], \nonumber \\
        \frac{\partial b}{\partial t} &=& \frac{2}{\langle k\rangle} \sum_{j\in \left\{ \uparrow,\downarrow\right\}} c_{j}
        \left\{ (1-p) \theta_{j}^{2q}
        \left[  \langle k\rangle-2q -2 \left( \langle k\rangle -q\right) \theta_{j} \right]
        +\frac{p}{2}\langle k\rangle \left(1-2 \theta_{j}\right) \right\}.
        \label{systemcb5vG}
    \end{eqnarray}
\end{widetext}
It can be easily verified that in the framework of the PA the above $q$-voter model with independence in a whole range of $p$, $0\le p\le1$,
is equivalent to the $2q$-voter model with independence on an
aggregate monoplex network with mean degree $2 \langle k\rangle$ being a superposition of the two layers of the MN.

It should be emphasized that under two main assumptions consisting in the use of homogeneous PA and constraining attention only to
stationary thermodynamic phases (fixed points) the systems of equations for the macroscopic concentrations $c$ and $b$, Eq.\ (\ref{systemcb5v}),
(\ref{systemcb5vG}) are obtained in which the only dependence on the degree distribution within layers of the MN is via the mean degree of nodes
$\langle k\rangle$. As a result, in theoretical description 
dependence is lost of such basic quantities as the order of the transition and the location of critical point or points on
the details of the degree distributions within layers $P(\left( k^{(A)}\right)$, $P\left(k^{(B)}\right)$, in particular on their heterogeneity.
This leads to substantial discrepancies between the predictions of the PA and results of MC simulations in the case of the $q$-voter model on
MNs with strongly heterogeneous layers. Nevertheless, the homogeneous PA in principle takes into account the degree heterogeneity of nodes,
cf.\ Eq.\ (\ref{rate1}), (\ref{rate2}). Thus, when going beyond the adiabatic limit and discussing, e.g., fluctuations of the macroscopic quantities
within the homogeneous PA, the obtained results can depend on the heterogeneity of the degree distributions
\cite{Peralta18a,Peralta18}; however, such investigation is beyond the scope of this paper.


\section{Results and discussion}

\subsection{Methods of analysis}

In this section results of MC simulations of the mentioned $q$-voter models with independence on MNs with two layers in the form of complex networks
are presented and compared with predictions of the analytic approaches based on the MFA and PA.
Both layers have identical degree distributions $P(k)$ and mean degrees $\langle k\rangle$.
In particular, layers in the form of RRG, ERG and SF networks with various power law exponents $\lambda$, 
covering both homo- and heterogeneous networks are considered.
The order parameter for the model is the magnetization $m$ related to the concentration $c$ of nodes with spins with orientation
up by $c=(1+m)/2$. 
Assumption that the $q$-lobbies are chosen without repetition imposes certain constraints on the parameters of the
networks forming the layers of the MNs on which the model can be studied.
In the case of RRGs the condition $k_{0}=\langle k\rangle \ge q$ is sufficient to fulfil this assumption. In the
case of SF networks the sufficient condition is $k_{\rm min}\ge q$ which imposes a constraint on the minimum value of the mean degree of layers,
$\langle k\rangle \ge \left(\lambda -1\right)q/\left(\lambda-2\right)$. In the case of ERGs nodes with degrees $k^{(A)}<q$ or $k^{(B)}<q$ are always
present in the MN. In order not to distort the degree distribution by excluding such nodes from MC simulations it is assumed that in this case 
the whole neighborhood within the respective layer forms the $q$-lobby. In order to minimize the effect of such nodes on the numerical results
only layers in the form of ERGs with $\langle k\rangle \gg q$ are considered.

In general, numerical evolution of the models under consideration revealed existence of two phases: the PM phase with $m=0$ and the FM phase with 
$m \ne 0$ with continuous or discontinuous transition from the former to the latter with decreasing parameter $p$.
The overall scenario of this transition for the models on all kinds of investigated MNs with a broad range of mean degrees of nodes, 
with both LOCAL\&AND and GLOBAL\&AND spin update rules is the same, 
only details such as the critical values of the control parameter $p$ differ. 

For small values of $q$, i.e., for $q=2,3$ in the case with LOCAL\&AND spin update rule 
and for $q=2$ in the case with GLOBAL\&AND spin update rule the transition from FM to PM phase with decreasing $p$ is continuous.
In order to determine the possible universality class of this transition in Sec.\ \ref{sec42} results of  
MC simulations were performed of the models on MNs with given topology of layers and different
numbers of nodes $N$ and with PM initial conditions with $m=0$. For a range of values of $p$ in the vicinity of the critical point time
series of  the instantaneous magnetization $\tilde{m} = N^{-1} \sum_{i=1}^N s_i$ were collected after initial transient. 
Next, the magnetization $m$, susceptibility $\chi$ and the fourth-order Binder cumulant $U_4$ were evaluated as
functions of $p$,
\begin{eqnarray}
m(p) & = & \left[ \langle | \tilde{m} | \rangle_{t} \right]_{av},
\label {M}\\
\chi(p) & = & N \left[\left(
\langle \tilde{m}^2 \rangle_{t} -  \langle \left| \tilde{m} \right| \rangle_{t} ^2
\right)\right]_{av},
\label{chi} \\
 U_{4}(p)& = &\frac{1}{2}\left[ 3-\frac{\langle \tilde{m}^{4} \rangle_{t}}{\langle \tilde{m}^{2}\rangle_{t}^{2}} \right]_{av},
\label{U4}
\end{eqnarray}
where $\langle \cdot \rangle_{t}$ denotes time average for a given realization of the MN network
and  $[\cdot]_{av}$ denotes averaging over different realizations of the MN network 
with given $N$ and degree distribution $P(k)$.  
The above quantities are expected to obey finite size scaling (FSS) relations analogous to those valid for
equilibrium systems on complex heterogeneous networks \cite{Hong07},
\begin{eqnarray}
m &=& N^{-\beta/\nu} f_m \left( N^{1/\nu}(p-p_c) \right) \label{fssM} \\
\chi &=& N^{\gamma/\nu} f_\chi \left( N^{1/\nu}(p-p_c) \right) \label{fsschi}\\
p_{c}-\tilde{p}(N)&\propto & N^{-1/\nu}, \label{fssqc}
\label{eq:scalling}
\end{eqnarray}
where $\tilde{p}(N)$ denotes the value of $p$ for which
the susceptibility $\chi$ of the model on a network with $N$ nodes has a maximum value.

The critical value $p_{c}$ for the FM transition is obtained from the intersection
point of the Binder cumulants for different sizes $N$ of the MN \cite{Binder97}. Next, from
Eq.\ (\ref{fssM},\ref{fsschi}) the exponents $\beta/\nu$ and $\gamma/\nu$, respectively, are
determined. Furthermore, Eq.\ (\ref{fssqc}) is used to calculate
the exponent $1/\nu$; as a result, the critical exponents $\beta$ and $\gamma$ are finally evaluated. 
Eventually, it is verified if the obtained exponents fulfil the hyperscaling relation,
\begin{equation}
2\frac{\beta}{\nu}+\frac{\gamma}{\nu}=D_{eff},
\label{Deff}
\end{equation}
where the effective dimension $D_{eff} = 1$ is expected
in the case of systems on MNs with layers in the form of complex networks
which do not have any particular spatial dimension \cite{Hong07}.

For $q>4$ in the case of the LOCAL\&AND spin update rule and for $q\ge 3$ in the case of the GLOBAL\&AND spin update rule the FM transition in the
$q$-voter model on different MNs becomes discontinuous, 
with hysteresis loop which size depends on the topology of the layers, particularly on $\langle k \rangle$.
In order to determine stability regions of the PM and FM phases and the width of the possible hysteresis loop 
simulations for each value of $p$ are run with two different sets of initial conditions: random PM set with $m=0$ and uniform FM set with $m=1$.
After long enough transient the time-averaged magnetization is evaluated based on a large number of MC steps following the transient.
From simulations the average magnetization $m$ and thus concentration $c$ in the resulting stationary state of the system are obtained, 
which are later compared with predictions of the homogeneous PA.

Analysis of the theoretical equations resulting from the PA presented in Sec.\ \ref{sec43} was performed using numerical methods of solving 
systems of algebraic equations
applied to Eq.\ (\ref{systemcb5v}) and (\ref{systemcb5vG}) in order to find their equilibria and to determine their stability,
as well as a Dormand-Prince order 4/5 Runge-Kutta method to obtain time evolution of the concentrations $c$, $b$.
Equilibria of the systems of equations (\ref{systemcb5v}) 
and (\ref{systemcb5vG}) are solutions of
equations $\partial b/\partial t =0$, $\partial c/\partial t=0$. Different stable equilibria correspond to different thermodynamic phases 
of the models under study. 
In general, numerical results and predictions of the PA show quantitative or at least qualitative (e.g., concerning the order of the FM transition)
agreement for the model on MNs with
the mean degree of nodes within layers $\langle k\rangle$ large enough, in particular for $\langle k\rangle \gg q$. Otherwise, for small
$\langle k\rangle$ comparable with $q$, numerical and theoretical results differ substantially. Thus, in Sec.\ \ref{sec43}  the two above-mentioned 
regimes are discussed separately.

\subsection{Finite size scaling and critical exponents}

\label{sec42}

\begin{figure}[b]
    \includegraphics[width=0.32\linewidth]{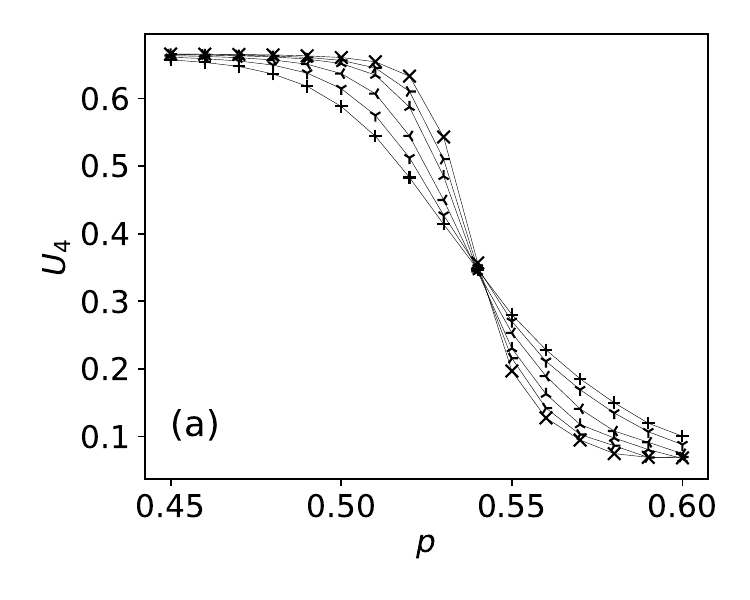}
    \includegraphics[width=0.32\linewidth]{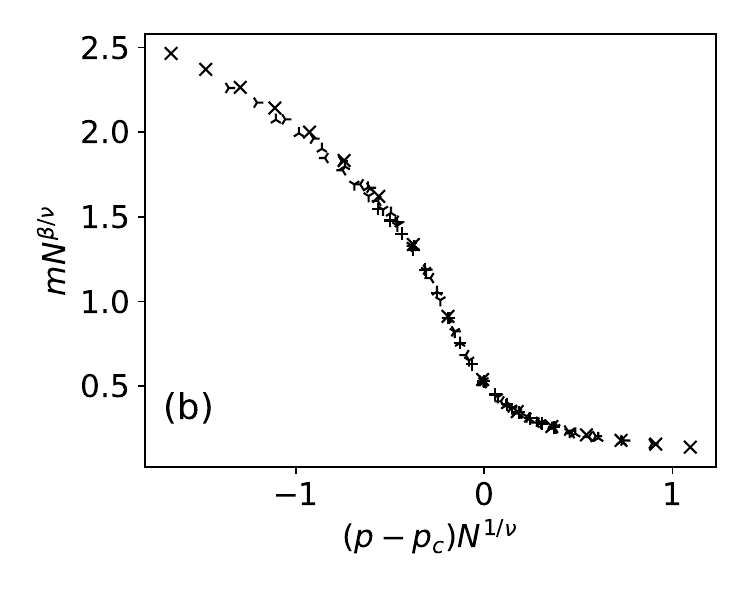}
    \includegraphics[width=0.32\linewidth]{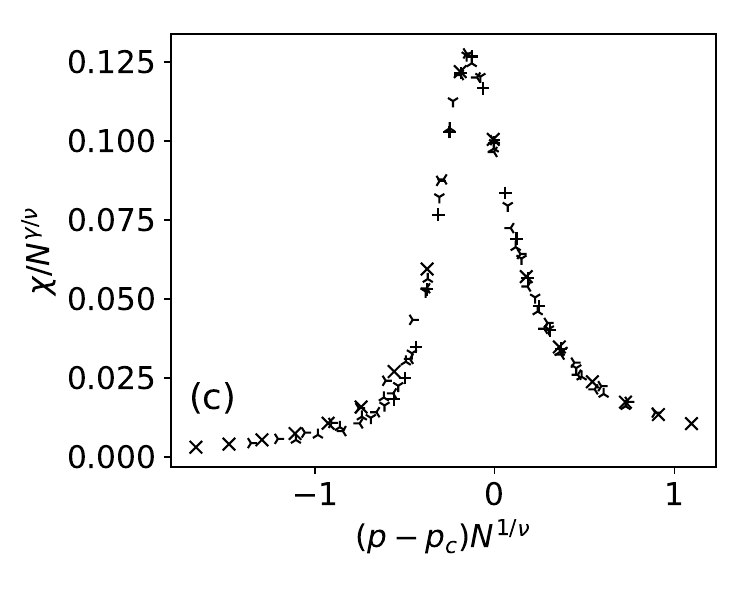}
    \caption{\label{fig:exp} Results of MC simulations of the $q$-voter model 
with independence with $q=4$ and LOCAL\&AND spin update rule on
a MN with SF layers with $\gamma=2.5$ and $k_{min}=10$. (a) Binder cumulants $U_4$ vs. $p$ for $N=$ 
    	$5 \cdot 10^2$ ($+$), $10^3$, $2 \cdot 10^3$, $5 \cdot 10^3$, $10^4$, $2 \cdot 10^4$ ($\times$), (b) magnetization $m$ and (c) susceptibility $\chi$ rescaled according to Eq. (\ref{fssM}) and Eq. (\ref{fsschi}) respectively with critical exponents from Table \ref{t:exp}.}
\end{figure}

\begin{table}[h]
\begin{tabular}{l c c c c r}
    & $p_c$ & $\beta/\nu$ & $\gamma/\nu$ & $1/\nu$ &  $D_{eff}$ \\
    \hline
    SF $\lambda=2.5$, $k_{min}=10$ & 0.541 & 0.129(9) & 0.748(5) & 0.279(28) & 1.006 \\
    SF $\lambda=3.0$, $k_{min}=10$ & 0.535 & 0.209(3) & 0.594(5) & 0.453(37) & 1.013 \\
    RRG $k=10$ & 0.5152 & 0.251(2) & 0.504(1) & 0.503(19) &  1.006 \\
\end{tabular}
   \caption{Critical value of the independence parameter $p_c$, critical exponents  $\beta/\nu$, $\gamma/\nu$, $1/\nu$ and effective dimension $D_{eff}$ of the $q$-voter model with independence with q=4 and LOCAL\&AND spin update rule on MNs with layers with different degree distributions.} 
   \label{t:exp}
\end{table}

In order to find the possible universality class of the FM transition for small $q$ the $q$-voter model on
MNs with large mean degrees $\langle k\rangle \gg q$ and different degree distributions of layers was investigated
numerically. Exemplary curves $m(p)$, $\chi(p)$, $U_4 (p)$
for the LOCAL\&AND spin update rule are shown in Fig. \ref{fig:exp}  and the results are summarized in Tab.\ {\ref{t:exp}}.
The estimated critical exponents turn out to be similar to those for the $q$-voter model on (monoplex) networks
with the corresponding degree distributions \cite{Peralta18}. In particular, in all cases the exponent $\beta$ is slightly
below $1/2$
which is the value following from the MFA in Sec.\ III.A using the Landau theory. On the other hand the remaining
critical exponents depend on the topology of connections. For homogeneous and weakly heterogeneous layers 
(RRGs, ERGs, SF networks with $\lambda >3$) the exponents are close to their MF values,
$\gamma =1$, $\nu =2$. However, for strongly heterogeneous layers (SF networks with $\lambda <3$) they seem to
be non-universal and depend on the degree distribution. In Ref.\ \cite{Peralta18} arguments were given to determine
this dependence in the case of the $q$-voter model on SF networks with $\lambda <3$.
In particular, it was shown that the exponent $\nu$ fulfils the
relation $1/\nu = (1-b)/2$, where $b$ is the scaling exponent for the dependence of the second moment of the degree
distribution on the number of nodes, $\langle k^2 \rangle =\int_{k_{\rm min}}^{k_{\rm max}} P(k)k^2 dk \propto N^b$,
which in turn can be estimated taking into account that the maximum degree $k_{\rm max}$ is approximately such that
probability to find a node with degree $k>k_{\rm max}$ is of order $1/N$, i.e., 
$\int_{k_{\rm max}}^{\infty} P(k) dk \approx N^{-1}$, and the degree distribution is normalized as
$\int_{k_{\rm min}}^{k_{\rm max}} P(k) dk =1$. This yields $k_{\rm max} \approx k_{\rm min}N^{1/(\lambda -1)}$,
$\langle k^2 \rangle \propto N^{(3-\lambda)/(\lambda -1)}$, i.e., $b= \frac{3-\lambda}{\lambda -1}$ and
$1/\nu =\frac{\lambda -2}{\lambda -1}$. For the $q$-voter model on a MN with SF layers with $\lambda =2.5$
from MC simulations $\nu = 3.59$ was obtained (Table \ref{t:exp}), which is comparable with
the predicted value $\nu=3.0$ and 
much above the values $\nu \approx 2$ obtained in the case of weakly heterogeneous layers. Finally, in all cases
under study the hyperscaling relation, Eq.\ (\ref{Deff}), is fulfilled with good accuracy.

\subsection{Comparison of predictions of the homogeneous pair approximation with numerical simulations}

\label{sec43}

\begin{figure}[b]
    \includegraphics[width=\linewidth]{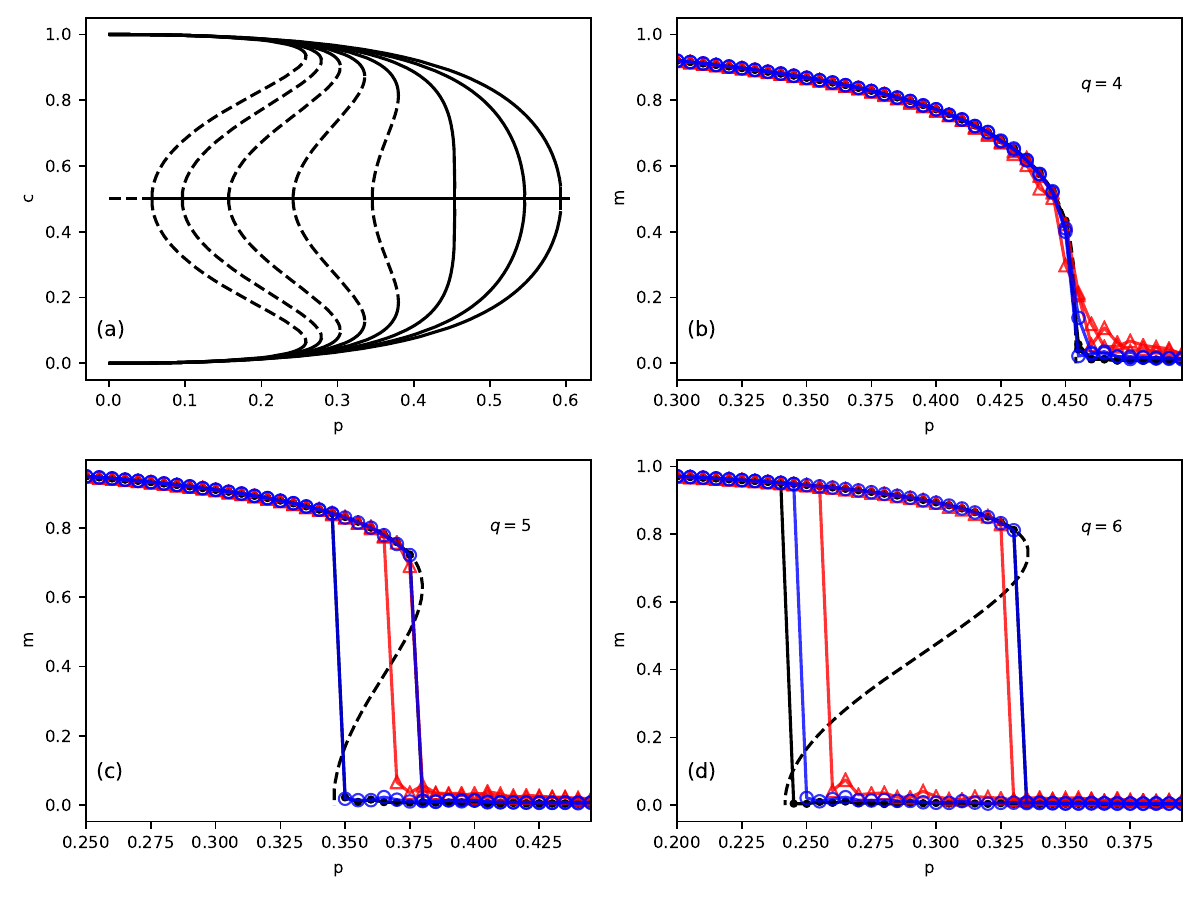}
    \caption{\label{fig:la} LOCAL\&AND rule on duplex network. (a): Concentration $c$ vs.\ degree of stochasticity parameter 
$p$ according to the PA for fixed value of $\langle k \rangle = 40$. Stable fixed points denoted with solid line, unstable fixed points - with dashed line ($q=2..9$ from right to left). (b-d) Magnetization $m$ vs.\ parameter $p$ - numerical results for the ERG (black dots), SF with $\lambda=3$ (blue circles) and SF with $\lambda=2.5$ 
    (red triangles) compared with the PA results for different values of $q$.}
\end{figure}

\begin{figure}[b]
    \includegraphics[width=\linewidth]{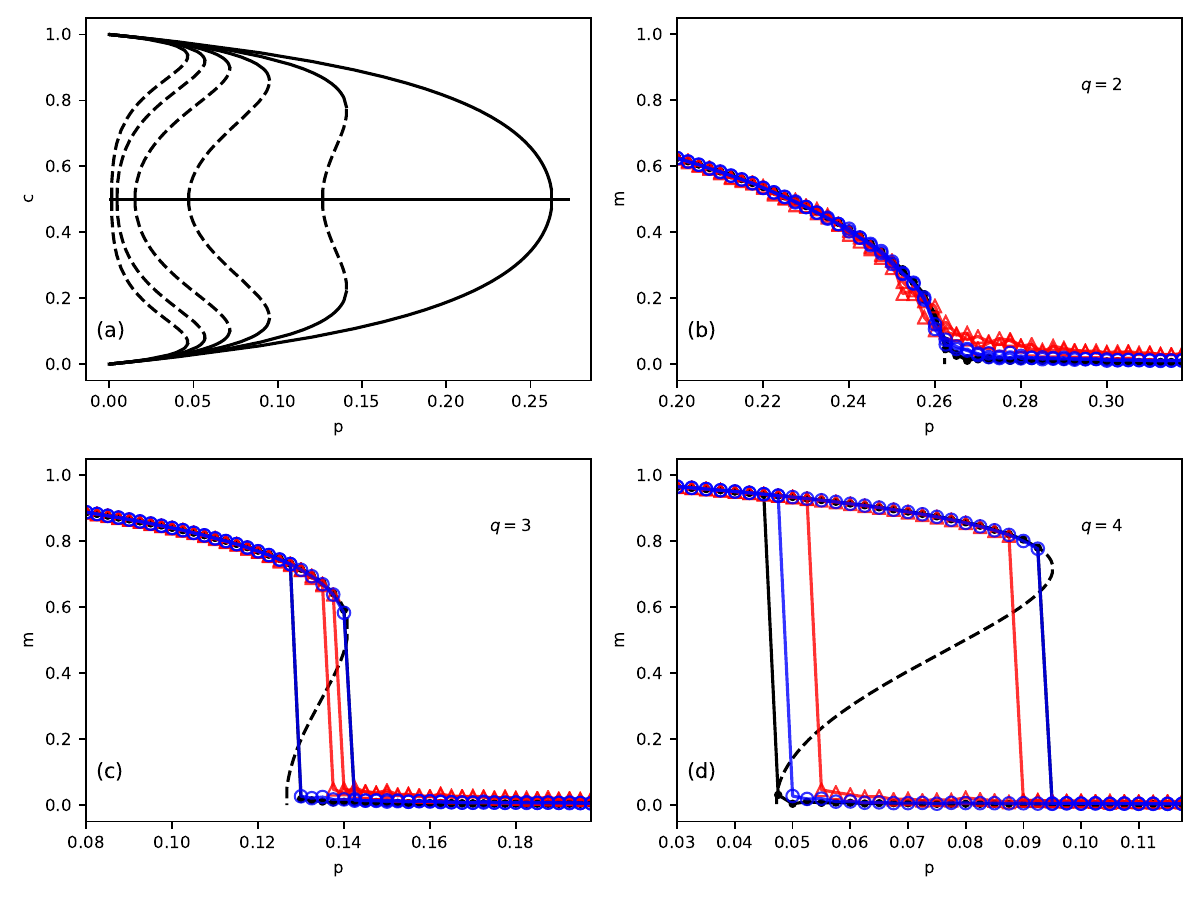}
    \caption{\label{fig:ga} GLOBA\&AND rule on duplex network. (a): Concentration $c$ vs.\  degree of stochasticity 
$p$ according to the PA for fixed value of $\langle k \rangle = 40$. Stable fixed points denoted with solid line, unstable fixed points - with dashed line ($q=2..7$ from right to left). (b-d) Magnetization $m$ vs.\ parameter $p$ - numerical results for the ERG (black dots), SF with $\lambda=3$ (blue circles) and SF with $\lambda=2.5$ 
    (red triangles) compared with the PA results for different values of $q$.}
\end{figure}

In this section comparison is performed between the predictions of the homogeneous PA and results of MC simulations of the $q$-voter model
on MNs with layers formed by various complex networks.
Let us start with the case with large $\langle k\rangle$, so that $\langle k\rangle \gg q$. First, results obtained from the homogeneous PA for this case are
outlined; their overall form is the same for both LOCAL\&AND and GLOBAL\&AND spin update rules.
Depending on the parameters $\langle k\rangle$, $q$ two different kinds of bifurcation diagrams of the systems 
of equations Eq.\ (\ref{systemcb5v}) and (\ref{systemcb5vG})
are obtained as the parameter $p$ is varied, typical of the continuous or discontinuous phase transition. 
In both cases at high $p$ the only stable fixed point is  $c=1/2$ ($m=0$), $b\le 1/2$, corresponding to the PM
phase. In the scenario corresponding to the continuous transition to the FM phase as $p$ is
decreased this point loses stability via a supercritical pitchfork bifurcation
at $p=p_{c}$ and for $p<p_{c}$ a pair of stable equilibria
with $ c >1/2$ ($m>0$), $b<1/2$, or $c<1/2$ ($m<0$), $b<1/2$, exists, corresponding to the FM phase with positive or negative magnetization,
respectively. In the scenario corresponding to the discontinuous transition as $p$ is
decreased two pairs of stable and unstable equilibria
appear via two saddle-node bifurcations taking place simultaneously at $p=p_{c}^{(2)}$. For
$p_{c}^{(1)}< p< p_{c}^{(2)}$ the two above-mentioned stable equilibria, one with 
$ c >1/2$ ($m>0$), $b<1/2$, and the other with $c<1/2$ ($m<0$), $b<1/2$, 
corresponding again to the FM phase with positive or negative magnetization, respectively,
coexist with the stable equilibrium with $c=1/2$ ($m=0$), $b \le 1/2$, corresponding to the PM phase;
the basins of attraction of the three stable equilibria are separated by stable manifolds of the two unstable equilibria.
Eventually at $p=p_{c}^{(1)}$ the fixed point corresponding to the PM phase loses stability via a subcritical
pitchfork bifurcation by colliding with the above-mentioned pair of unstable equilibria, and for $p<p_{c}^{(1)}$
the only two stable fixed points are those corresponding to the FM phase. Hence, $p_{c}^{(1)}$, $p_{c}^{(2)}$ correspond
to the lower and upper critical value of $p$ for the first-order transition, respectively, and
for $p_{c}^{(1)}< p< p_{c}^{(2)}$ stable PM and FM phases coexist. It should be noted that the values of 
$p_{c}$ for the continuous and $p_{c}^{(1)}$ for the discontinuous transition to the FM phase can be
obtained analytically by means of linear stability analysis of the PM fixed point of Eq.\ (\ref{systemcb5v}) or 
(\ref{systemcb5vG}) with $c=1/2$ ($m=0$) (see Appendix, Eq.\ (\ref{pstarLA1}) and (\ref{pstarGA1})).

\begin{figure}[b]
   \includegraphics[width=0.49\linewidth]{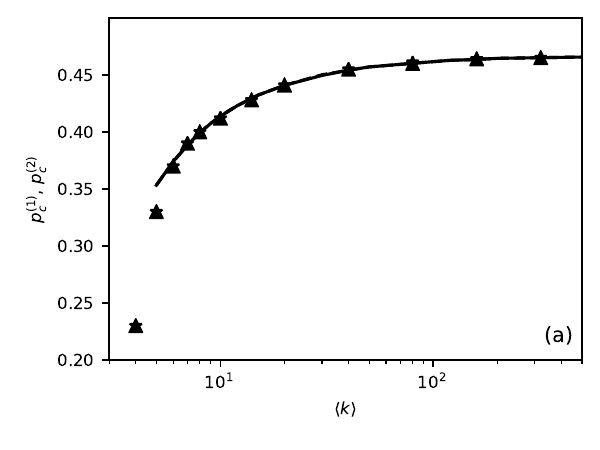}
   \includegraphics[width=0.49\linewidth]{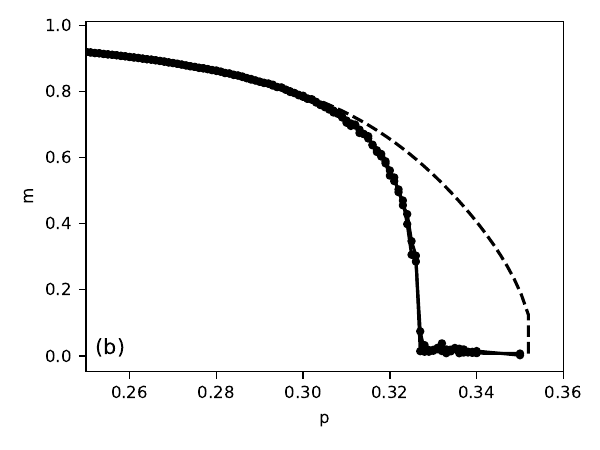}
   \includegraphics[width=0.49\linewidth]{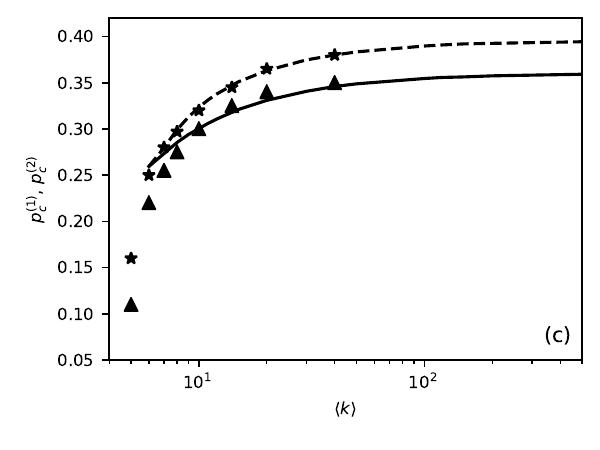}
   \includegraphics[width=0.49\linewidth]{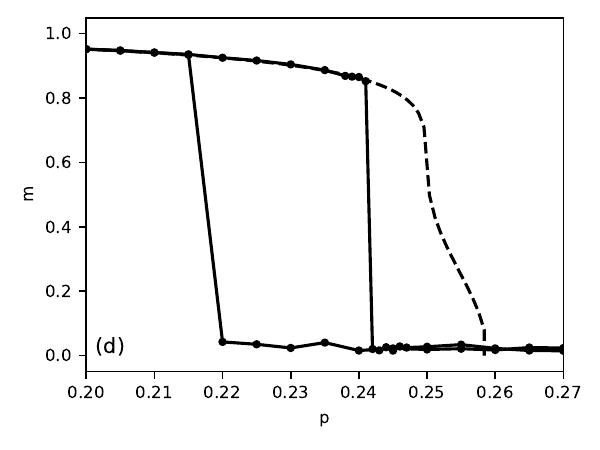}
   \includegraphics[width=0.49\linewidth]{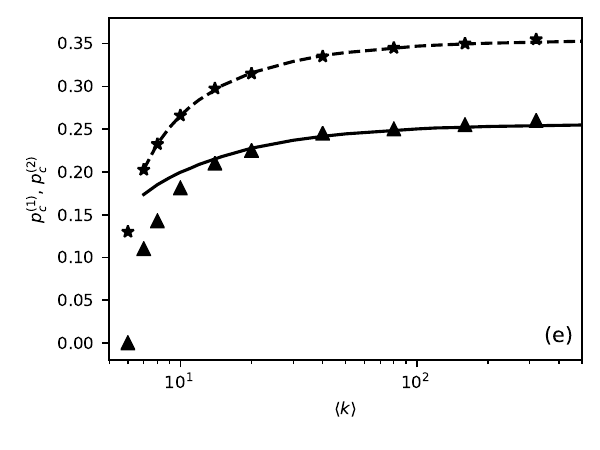}
   \includegraphics[width=0.49\linewidth]{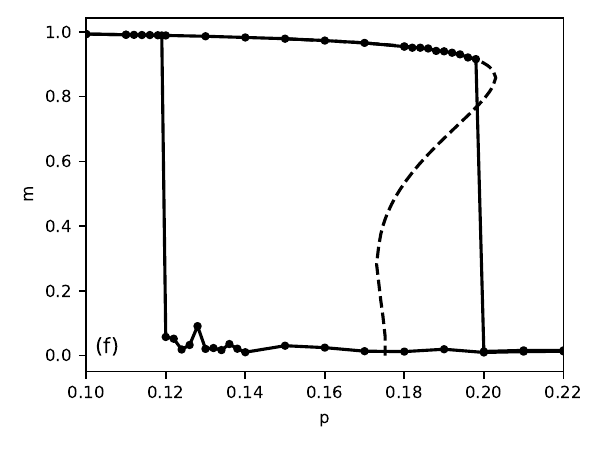}
   
    \caption{\label{fig:pc} Comparison between the lower ($p_c^{(1)}$) and upper ($p_c^{(2)}$) critical values of parameter $p$ for the LOCAL\&AND rule model with (a) $q=4$, (c) $q=5$, (e) $q=6$, obtained from MC simulations of the model on RRGs (symbols) and from PA (lines). Magnetization $m$ vs.\ parameter $p$ - numerical results for the RRG (connected
black dots) compared with the PA results for: (b) $q=4$, $\langle k\rangle =5$, (d) $q=5$, $\langle k\rangle =6$,
(e) $q=6$, $\langle k\rangle =7$.}
\end{figure}

Provided that $\langle k\rangle\gg q$, according to the homogeneous PA 
the models under consideration exhibit first-order phase transition to FM phase for $q \ge 5$ in the case of LOCAL\&AND spin update rule and for $q \ge 3$ 
for the GLOBAL\&AND rule; otherwise, the transition is second-order.
This is illustrated in Fig.\ \ref{fig:la}(a) and Fig.\ \ref{fig:ga}(a), respectively, for the case of the model on MNs with layers with relatively large 
mean degree of nodes $\langle k\rangle$. 
The above-mentioned minimal values of the size of the $q$-lobby
for the occurrence of the discontinuous FM transition agree with those in the appropriate models
on MNs with layers in the form of complete graphs \cite{Chmiel15} which were analytically confirmed by the MFA (Sec.\ III A). In the case of 
LOCAL\&AND spin update rule it is difficult to verify analytically if the PA predicts occurrence of the tricritical point 
separating the first- and second-order transition exactly at $q=4$, nevertheless,
numerical analysis of the stable fixed points of Eq.\ (\ref{systemcb5v}) supports this conjecture (Fig.\ \ref{fig:la}(a)). In contrast, in the case of the
GLOBAL\&AND update rule there is no numerical evidence for the existence of any tricritical point (Fig.\ \ref{fig:ga}(a)), again in  agreement with predictions of the MFA. 

Although in the framework of the homogeneous PA the critical values of the parameter $p$ as well as the location of the fixed points
corresponding to the PM and FM phases depends only on the mean degree $\langle k\rangle$,
 MC simulations reveal that
the details of the continuous or discontinuous FM transition depend also 
on the topology of the layers rather than only on the first moment of the degree distribution.
Comparison of the theoretical approach with numerical simulations of the model on MNs with various topologies of the layers, presented in Fig.\ \ref{fig:la} (b-d) 
and Fig.\ \ref{fig:ga} (b-d), shows very good quantitative agreement between theory and simulations for layers in the form of homogeneous networks, 
in particular ERG, RRG (results not shown on plots due to almost full overlap with 
those for ERG) and weakly heterogeneous SF networks with exponent $\lambda \ge 3$
(i.e., with finite second moment of the degree distribution).
Both the order of the transition and the width of the possible hysteresis loop are predicted correctly by the PA.
In contrast, numerical results for MNs with layers in the form of strongly heterogeneous SF networks with exponent $2<\lambda <3$ significantly differ from the
theoretical predictions, which is illustrated in  in Fig.\ \ref{fig:la} (b-d) and Fig.\ \ref{fig:ga} (b-d) for $\lambda=2.5$. 
The discrepancies become even bigger for smaller values of $\lambda$ (not shown on plots) which manifests itself with shrinking, 
or even practically disappearing, hysteresis loop. Nevertheless, despite quantitative differences, the order of the phase transition is predicted correctly by the 
homogeneous PA, thus there is qualitative agreement between theoretical and numerical results.

In the case of MNs with homogeneous or weakly heterogeneous layers 
agreement between numerical and theoretical results based on the 
homogeneous PA is very good in a broad range of the mean degrees of nodes $\langle k \rangle$
provided that $\langle k\rangle \gg q$.
This is illustrated in Fig. \ref{fig:pc}(a, c, e) where location of the critical point or points in the case of second- and first-order FM transition, respectively, is shown 
as a function of $\langle k\rangle$ for the model with LOCAL\&AND spin update rule with different sizes of the $q$-lobby. 
The position of the critical point $p_{c}$ in the case of the second-order transition (Fig.\ 3(a))
and of the critical points $p_{c}^{(1)}$, $p_{c}^{(2)}$ in the case of the first-order transition (Fig.\ 3(c, e))
is very well predicted directly from simulations of Eq.\ (\ref{systemcb5v}). 
In the case of the critical point $p_{c}$ for the second-order transition and the lower critical point $p_c^{(1)}$ for the first-order transition
this prediction coincides with the analytic result of Eq.\ (\ref{pstarLA1}) in the Appendix.
The values of $p_{c}$ or $p_{c}^{(1)}$ and $p_{c}^{(2)}$ as well as the width of the possible hysteresis loop increase with $\langle k\rangle$ 
and saturate at maximum values for $\langle k \rangle \to \infty$, with the topology of the layers approaching that of a complete graph.
This proves that the homogeneous PA captures all details and provides quantitatively correct description of the FM transition in 
the $q$-voter model with independence on MNs with homogeneous or weakly heterogeneous layers with large enough mean degrees of nodes within layers.
This description is significantly improved in comparison with that based on the MFA which results do not depend even on such
basic feature of the layers as the mean degree of nodes \cite{Chmiel15}. In contrast, as mentioned above, in the case of MNs with strongly
heterogeneous layers the homogeneous PA, which does not assume any particular topology of the network, is insufficient to describe quantitatively
the details of the FM transition in the $q$-voter model under study.

As mentioned in Sec.\ III.C in the framework of PA the $q$-voter model with independence on MNs 
with the GLOBAL\&AND spin update rule is equivalent to the 
$2q$-voter model on an aggregate monoplex network being a superposition of the two layers.
This equivalence has already been observed in the $q$-voter model on MNs with layers in the form of complete graphs \cite{Chmiel15}. 
Performed MC simulations of the $q$-voter model on duplex networks also confirm that it is fully equivalent
to the $q$-voter model on appropriate aggregate monoplex networks, independently of the topology and for a broad range 
of the mean degree of nodes in layers (not shown). 
Therefore, small differences in the dynamics of these two models mentioned in Sec.\ II.A
do not have visible effect on the observed phase transitions.

Let us in turn consider the models on MNs with layers with the mean degree of nodes $\langle k\rangle$ small and comparable with the
size of the $q$-lobby. In this case predictions of the homogeneous PA and results of MC simulations differ substantially even for
the models on MNs with layers in the form of homogeneous ERG or RRG. These differences are illustrated in Fig.\ 3 for the model with
the LOCAL\&AND spin  update rule. Again, MC simulations reveal that for $q\le4$ the model undergoes continuous (Fig.\ 3(a,b))
and for $q\ge 5$ discontinuous FM transition  (Fig.\ 3(c-f)). For $q=4$ and small $\langle k\rangle$ the PA correctly predicts the
occurrence of the continuous FM transition, but the predicted critical value $p_{c}$ significantly exceeds that obtained numerically 
(Fig.\ 3(a,b)), in contrast with what is observed for larger $\langle k\rangle$. For $q=5$ both the lower and upper critical values
$p_{c}^{(1)}$, $p_{c}^{(2)}$ obtained from the MC simulations decrease fast with decreasing $\langle k\rangle$, and for $\langle k\rangle <10$
the numerical value of $p_{c}^{(1)}$ becomes much lower than that predicted by the PA (Fig.\ 3(c)). Nevertheless, for $\langle k\rangle >6$
the theory correctly predicts that the transition is first-order, and 
the numerical value of $p_{c}^{(2)}$ agrees with that predicted from the PA (Fig.\ 3(c)). 
However, for $\langle k \rangle \le 6$ the PA incorrectly predicts that the lower and upper critical 
values $p_{c}^{(1)}$, $p_{c}^{(2)}$ merge, the transition is second-order and occurs at higher $p_{c}$ than observed in simulations (Fig.\ 3(c,d)).

Discrepancies between the theory based on the PA and results of the MC simulations are even more significant for
$q=6$ (Fig.\ 3(e,f)). For example, for $\langle k\rangle =7$ the PA predicts that with decreasing $p$ the PM fixed point at $m=0$
loses stability at $p=p_{c}^{(1)}$, but in a narrow interval of the parameter $p$ just below $p_{c}^{(1)}$ another stable fixed point  appears
corresponding to a phase with partial FM ordering characterized by small but non-zero value of the magnetization. Only as $p$ is further decreased
the latter fixed point loses stability and a discontinuous transition to the usual FM phase with $m\approx 1$ occurs (Fig.\ 3(f)). This complex scenario is not 
confirmed by the MC simulations where usual discontinuous jump of the magnetization is observed with decreasing $p$,
typical of the first-order FM transition to a highly ordered phase, at a critical value $p_{c}^{(1)}$ much lower than that predicted theoretically
(Fig.\ 3(f)). In contrast, for increasing $p$ the theory correctly predicts that the transition from the FM to the PM phase is discontinuous
and the predicted upper critical value $p_{c}^{(2)}$ agrees well with that obtained from the MC simulations (Fig.\ 3(e,f)). For $q\ge 7$ 
and small $\langle k\rangle$ such substantial differences between the numerical and theoretical scenarios for the FM transition in the model 
under study do not occur. Nevertheless, the examples above show that the theory based on the homogeneous PA, though more elaborate than that based on a simple
MFA, can also fail for the $q$-voter model with independence 
on MNs with homogeneous  layers with low density of edges (similar observation was reported for the case of the
$q$-voter model on monoplex networks \cite{Jedrzejewski17}).

\begin{figure}[b]
    \includegraphics[width=0.32\linewidth]{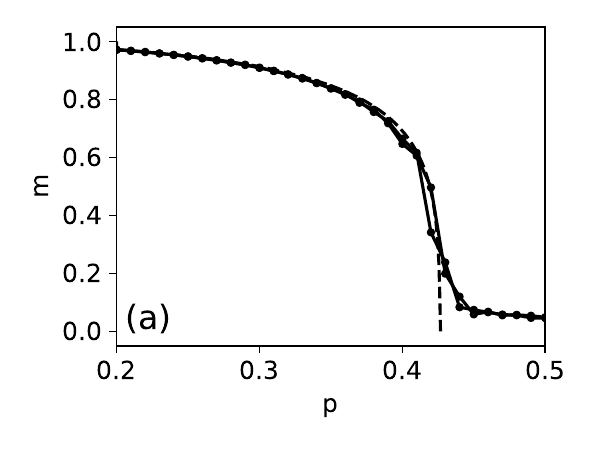}
    \includegraphics[width=0.32\linewidth]{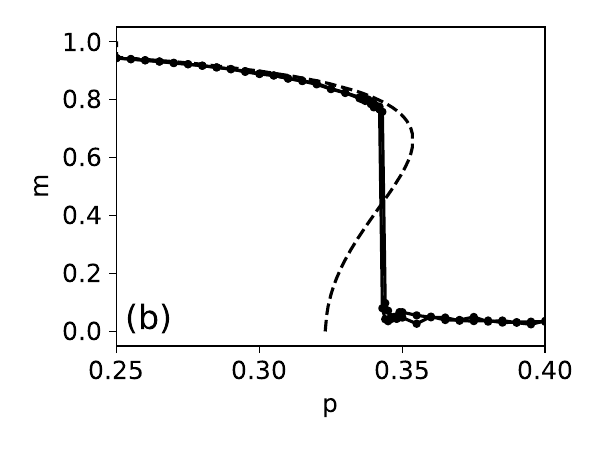}
    \includegraphics[width=0.32\linewidth]{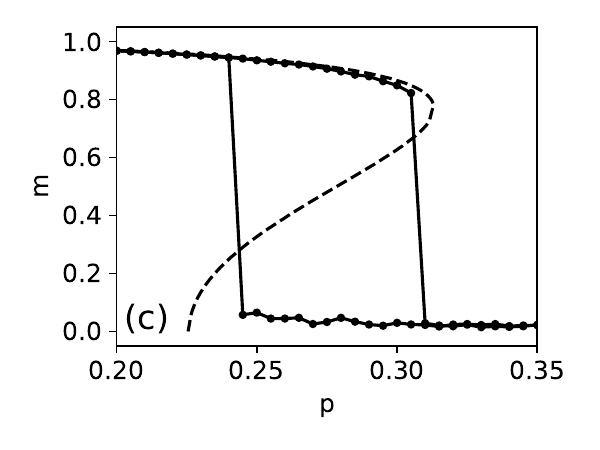}

    \caption{\label{fig:SFk_small} 
    		Magnetization $m$ vs.\ parameter $p$ for the $q$-voter model with the 
LOCAL\&AND spin update rule on a MN with SF layers - numerical results for $\lambda=2.5$ (connected
    		black dots) compared with predictions of the homogeneous PA
 for: (a) $q=4$, $k_{min}=5$ (b) $q=5$, $k_{min}=6$ (c) $q=6$, $k_{min}=7$.
    }
\end{figure}

For completeness, the $q$-voter model with different sizes of the $q$-lobby 
was also studied on MNs with strongly heterogeneous layers and the smallest possible value $k_{\rm min} =q+1$
to assure that the agents forming the $q$-lobby can be chosen without repetitions for all nodes. The dependence
of the magnetization $m$ on the parameter $p$ is shown in Fig.\ \ref{fig:SFk_small} and compared with predictions of the
homogeneous PA for the model with the LOCAL\&AND
spin update rule and small values of $q$ for which second- or first-order FM transition occurs. In all cases differences
between the obtained numerical results and those for
 the model with the same $q$ and large $\langle k\rangle$ are only quantitative (cf.\ Fig.\ \ref{fig:la}),
e.g., the width
of the hysteresis loop is smaller or even unnoticeable as in the $q=5$, $k_{\rm min}=6$ case (Fig. \ref{fig:SFk_small}(b))
but the
order of the transition is not changed. Also predictions of the homogeneous PA, though quantitatively incorrect, do not show
any dramatic qualitative discrepancies with the results of MC simulations. This is probably due to the fact that even for
the smallest possible values of $k_{\rm min}$ the mean degrees of nodes within layers $\langle k\rangle$ remain relatively
large, e.g., for $\lambda =2.5$ $\langle k\rangle \ge 3q$ is expected (Sec.\ IV.A) which improves the accuracy
of the theoretical analysis.


\section{Summary and conclusions}

In this paper the $q$-voter model with independence, which is a sort of model for the opinion formation, was studied on MNs with two
layers in the form of complex networks with identical degree distributions and full overlap of nodes, corresponding to 
different levels of social influence. The presence of the two layers was taken into account by assuming the LOCAL\&AND or GLOBAL\&AND
rules for the opinion update of the agents, which differ by the way in which the lobby of neighbors influencing the opinion of a given agent is formed. 
Both in theoretical investigations based on the MFA and homogeneous PA as well as in MC simulations FM phase
transition was observed as the parameter $p$ controlling the level of agents' independence was changed. This transition
can be first- or second-order, depending on the size of the $q$-lobby, and the details of the transition , 
e.g., location of the critical points, critical exponents and the width of the possible hysteresis loop, 
depend on the topology of the underlying MN. 
The homogeneous PA was derived for a more general case of a two-state spin model with up-down symmetry on MNs and applied 
to the above-mentioned $q$-voter model. Good agreement was obtained between predictions of this PA and results of MC simulations
for the model on MNs with layers with moderate and large mean degrees of nodes $\langle k\rangle$, in particular significantly larger than the size of the 
$q$-lobby. Then theoretical and numerical results
show good quantitative agreement for the model on MNs with layers in the form of homogeneous
ERGs and RRGs and weakly heterogeneous SF layers. For the model on MNs with strongly heterogeneous SF layers this agreement is only
qualitative, e.g., the order of the transition is predicted correctly. Theoretical and numerical results diverge and can differ even qualitatively
for mean degrees of nodes $\langle k\rangle$ small and comparable with $q$, even for small degree of heterogeneity of the layers forming
the MN. Results obtained in this paper can be easily extended to other models, e.g.,
the $q$-neighbor Ising model, as well as to models on MNs with more than two layers and layers with different topologies.

It can be seen that the systems of equations (\ref{systemcb5v}) and (\ref{systemcb5vG})
for the concentrations $c$, $b$ obtained in the homogeneous PA depend only on the 
mean degree of nodes $\langle k \rangle$ within each layer of the MN rather than on the precise form of the degree distribution $P(k)$.
This is probably the main source of quantitative discrepancy between theoretical and numerical results for the 
width and location of the hysteresis loop in the case of the first-order transition to the FM phase in the $q$-voter model with 
independence on MNs with strongly heterogeneous SF layers with $\lambda <3$. It can be expected that predictions based on some form of
heterogeneous PA will yield better agreement with results of MC simulations. Such predictions can be obtained, e.g., by
extending the general formulation of the heterogeneous PA for systems on monoplex networks \cite{Gleeson11,Gleeson13} to the case of MNs,
as it was done in Ref.\ \cite{Choi19} for the majority-vote model on MNs.
It should be mentioned that also the cases of the $q$-voter model with independence on MNs with partial
overlap of nodes and with correlations between degrees of nodes within different layers can be relatively easily studied in the framework of 
the above-mentioned extension.


\section*{Appendix}

In this Appendix stability of the PM phase for the $q$-voter model with independence on a MN with two layers with identical degree distributions
is investigated. For the LOCAL\&AND and GLOBAL\&AND spin update rules the PM phase corresponds to the fixed point of the two-dimensional systems of
equations, Eq.\ (\ref{systemcb5v}) and Eq.\ (\ref{systemcb5vG}), 
respectively, characterized by $c=1-c=1/2$. By performing linear stability analysis
it is shown that with decreasing $p$ the PM fixed point loses stability at
$p=p^{\star}$ ($0< p^{\star} <1$); in the case of the second-order transition to the FM phase $p^{\star}$ 
corresponds to the critical value $p_{c}$, while in the case of the first-order transition it corresponds to the lower critical value $p_{c}^{(1)}$. 

Let us start with the LOCAL\&AND spin update rule and denote the right-hand sides of the system of equations (\ref{systemcb5v}) as
\begin{widetext}
    \begin{eqnarray}
        A(c,b)&=&(1-c) \left[ (1-p)\theta_{\downarrow}^{q} +\frac{p}{2}\right]^2 -
        c\left[ (1-p)\theta_{\uparrow}^{q} +\frac{p}{2}\right]^2 \label{Acb} \nonumber \\
        B(c,b)
        &=& \frac{2}{\langle k\rangle} \sum_{j\in \left\{ \uparrow,\downarrow\right\}} c_{j} 
        \left\{ (1-p) \theta_{j}^{q}
        \left[  \langle k\rangle-2q -2 \left( \langle k\rangle -q\right) \theta_{j} \right]
        +\frac{p}{2}\langle k\rangle \left(1-2 \theta_{j}\right) \right\}
        \left[ (1-p) \theta_{j}^{q} +\frac{p}{2}\right]. \label{Bcb}
    \end{eqnarray}
\end{widetext}
At fixed points there is $A(c,b)=0$, $B(c,b)=0$. The (stable or unstable) PM fixed point exists for a whole range of
the parameter $p$, i.e., for $0 \le p\le1$. Since at this point $c=1/2$, from the condition $A(c,b)=0$ follows that 
$\theta_{\uparrow}=\theta_{\downarrow}\equiv \theta$, where $0\le \theta \le 1/2$ depends on $p$, i.e., the
position of the PM fixed point is $\left( c=1/2, b=\theta\right)$.
Then, since for $0< p\le 1$ there is $(1-p) \theta^{q} +\frac{p}{2}>0$, for $c=1/2$ from the condition $B(c,b)=0$ follows that
\begin{equation}
(1-p) \theta^{q}
\left[  \langle k\rangle-2q -2 \left( \langle k\rangle -q\right) \theta \right] =
- \frac{p}{2}\langle k\rangle \left(1-2 \theta\right).
\label{Bcb0}
\end{equation}
Solution of this nonlinear equation yields the value of $\theta$ at the PM fixed point.

Stability of the PM fixed point can be determined from the eigenvalues of the Jacobian matrix of the right-hand sides of Eq.\ (\ref{Acb}) 
at the fixed point. For this purpose let us first evaluate
\begin{eqnarray}
\left. \frac{\partial \theta_{\downarrow}^{q}}{\partial c}\right|_{c=\frac{1}{2},b=\theta} &=& 
\left. \frac{q \theta_{\downarrow}^{q}}{1-c}\right|_{c=\frac{1}{2},b=\theta} =2q\theta^{q}, \nonumber\\
\left. \frac{\partial \theta_{\downarrow}^{q}}{\partial b}\right|_{c=\frac{1}{2},b=\theta} &=& 
\left. \frac{q \theta_{\downarrow}^{q-1}}{2(1-c)}\right|_{c=\frac{1}{2},b=\theta}= q\theta^{q-1}, \nonumber\\
\left. \frac{\partial \theta_{\uparrow}^{q}}{\partial c}\right|_{c=\frac{1}{2},b=\theta} &=& 
\left. - \frac{q \theta_{\uparrow}^{q}}{c}\right|_{c=\frac{1}{2},b=\theta}  = - 2q\theta^{q}, \nonumber\\
\left. \frac{\partial \theta_{\uparrow}^{q}}{\partial b}\right|_{c=\frac{1}{2},b=\theta} &=&  
\left. \frac{q \theta_{\uparrow}^{q-1}}{2c}\right|_{c=\frac{1}{2},b=\theta} = q\theta^{q-1}.
\end{eqnarray}
Using the above formulae it is easily obtained that
\begin{equation}
\left. \frac{\partial A}{\partial b} \right|_{c=\frac{1}{2},b=\theta} =0.
\end{equation}
Thus, the Jacobian at the PM fixed point has an overall form
\begin{equation}
\left|
\begin{tabular}{cc}
$\left. \frac{\partial A}{\partial c} \right|_{c=\frac{1}{2},b=\theta}$  & $0$ \\
$\left. \frac{\partial B}{\partial c} \right|_{c=\frac{1}{2},b=\theta}$  & 
$\left. \frac{\partial B}{\partial b} \right|_{c=\frac{1}{2},b=\theta}$ 
\end{tabular}
\right|,
\end{equation}
and its eigenvalues are $\lambda_{1}=\left. \frac{\partial A}{\partial c} \right|_{c=\frac{1}{2},b=\theta}$, 
$\lambda_{2}=\left. \frac{\partial B}{\partial b} \right|_{c=\frac{1}{2},b=\theta}$.
The PM fixed point loses stability as with decreasing $p$ one of the eigenvalues changes sign from negative to positive.
After some transformations it is obtained that
\begin{equation}
\lambda_{1}=
 \left[ (1-p) \theta^{q} +\frac{p}{2}\right]\left[ 2(1-p)(2q-1)\theta^{q}-p\right],
\end{equation}
and, taking into account Eq.(\ref{Bcb0}),
\begin{widetext}
    \begin{eqnarray}
        \lambda_{2}&=&
        \frac{2}{\langle k\rangle}\left[ (1-p) \theta^{q} +\frac{p}{2}\right] 
        \left\{
            (1-p)q\theta^{q-1}\left[ \langle k\rangle -2q-2( \langle k\rangle -q) \theta\right] 
     - 2(1-p)\theta^{q}\left( \langle k\rangle -q\right) -p\langle k\rangle \right\}. \label{lambda2}
        \end{eqnarray}
 \end{widetext}
 Concerning $\lambda_{2}$, multiplying both sides of Eq.\ (\ref{Bcb0}) by $q$ and dividing by $\theta$ and then inserting the result
in Eq.\ (\ref{lambda2}) it is finally obtained that
\begin{widetext}
\begin{eqnarray}
\lambda_{2}&=&
-\frac{2}{\langle k\rangle}\left[ (1-p) \theta^{q} +\frac{p}{2}\right] 
\left[
\frac{pq\langle k \rangle}{\theta}(1-2\theta)
+2(1-p)\theta^{q}\left( \langle k\rangle -q\right) +p\langle k\rangle 
\right].
\end{eqnarray}
\end{widetext}
Thus, for $0< p<1$, $0\le \theta \le 1/2$, $q\le \langle k\rangle$ there is $\lambda_{2}<0$. Hence, the PM fixed point
with decreasing $p$ loses stability when $\lambda_{1}$ crosses zero. 
Since $(1-p) \theta^{q} +\frac{p}{2}>0$ this happens at $p=p^{\star}$ such that
\begin{equation}
2\left(1-p^{\star}\right)(2q-1)\theta^{\star q}-p^{\star}=0,
\end{equation}
i.e., for
\begin{equation}
p^{\star}=\frac{2(2q-1)\theta^{\star q}}{1+2(2q-1)\theta^{\star q}},
\label{pstarLA1}
\end{equation}
where $\theta^{\star}$ denotes the value of $\theta$ at the critical point. 
Substituting $p=p^{\star}$ in Eq.\ (\ref{Bcb0}) yields finally
\begin{equation}
\theta^{\star}=\frac{\langle k\rangle -1}{2\langle k\rangle -1}.
\label{thetastarLA1}
\end{equation}
For $\langle k\rangle \rightarrow \infty$, i.e, for the case of layers in the form of fully connected graphs, there is
$\theta^{\star} \rightarrow 1/2$ and $p^{\star}= \frac{2q-1}{2q-1+2^{q-1}}$, 
in agreement with the MFA result, Eq.\ (\ref{pstarMFLA}).

In the case of the GLOBAL\&AND spin update rule the stability analysis of the PM fixed point of the system of equations (\ref{systemcb5vG})
can be performed in a similar way, which yields at the critical point
\begin{eqnarray}
p^{\star}&=&\frac{2(2q-1)\theta^{\star 2q}}{1+2(2q-1)\theta^{\star 2q}}, \label{pstarGA1}\\
\theta^{\star}&=&\frac{\langle k\rangle -1}{2\langle k\rangle -1}.  
\label{thetastarGA1}
\end{eqnarray}
The same result can be obtained by replacing $\langle k\rangle \rightarrow 2\langle k\rangle$,
$q\rightarrow 2q$ in Eq.\ (34) of Ref.\ \cite{Jedrzejewski17}, which yields the value of $p^{\star}$ for the $q$-voter model 
with independence on a 
monoplex network. This is in agreement with the fact that the system of equations
(\ref{systemcb5vG}) describing the $q$-voter model with GLOBAL\&AND spin update rule on a MN with two layers can be obtained from that
describing the $q$-voter model on a monoplex network by making the above-mentioned replacement (Sec.\ III.B).
Again, for $\langle k\rangle \rightarrow \infty$ there is
$\theta^{\star} \rightarrow 1/2$ and $p^{\star}= \frac{2q-1}{2q-1+2^{2q-1}}$,
in agreement with the MFA result,
Eq.\ (\ref{pstarMFGA}).

It should be noted that in Ref.\ \cite{Gleeson13} was shown that for any model with up-down symmetry on a (monoplex) RRG with degree $k$ the PA at the
critical point yields $\theta^{\star}= \frac{k-2}{2(k-1)}$. Eq.\ (\ref{thetastarLA1}) and (\ref{thetastarGA1})
show that this result is valid also for the $q$-voter model on MNs with two
independently generated layers in the form of RRGs with identical degree distributions 
provided that $k$ is replaced with $2k$, and only the values of $p^{\star}$, Eq.\  (\ref{pstarLA1}) and (\ref{pstarGA1}),
depend on the details of the LOCAL\&AND and GLOBAL\&AND spin update rules. In the latter case for MNs with layers in the form of complex networks 
in  Eq.\ (\ref{thetastarLA1}) and (\ref{thetastarGA1}) $k$ can be eventually replaced by $\langle k \rangle$, but this replacement is valid in the homogeneous PA only.

\end{document}